\documentclass[preprint,12pt]{article}
\usepackage{natbib}
\usepackage{epsfig, amssymb,amsfonts}
\usepackage{graphicx,color}
\usepackage[FIGTOPCAP]{subfigure}
\usepackage{amsmath}
\usepackage{multirow}
\usepackage{url}

\newtheorem{thm}{Theorem}
\newtheorem{prop}{Proposition}
\newtheorem{lem}{Lemma}

\newtheorem{rem}{Remark}

\def\tr{{\rm tr}}
\def\inv{^{-1}}

\def\vec{{\rm vec}}

\begin{document}
\title{Fast Bivariate P-splines: the Sandwich Smoother}
\author{Luo Xiao\thanks{Graduate student, Department of Statistical Science, Malott Hall,
Cornell University, New York 14853 (email: \texttt{lx42@cornell.edu}).},\
Yingxing Li\thanks{ Graduate student, Department of Statistical Science, Malott Hall,
Cornell University, New York 14853 (email: \texttt{yl377@cornell.edu}).},\
and David Ruppert\thanks{Andrew Schultz, Jr., Professor of Engineering, School of
Operational Research and Information Engineering,  Comstock Hall, Cornell University,
New York 14853 (email: \texttt{dr24@cornell.edu}). } }
\date{July 13, 2012}
\maketitle

\begin{abstract}
We propose a fast penalized spline method for bivariate smoothing. Univariate P-spline smoothers
(Eilers and Marx, 1996) are applied simultaneously along both coordinates. The new smoother has a sandwich form which suggested the name ``sandwich smoother'' to a referee. The sandwich  smoother has a tensor product structure that simplifies an asymptotic analysis and  it can be fast computed. We derive a local central limit theorem for the sandwich smoother, with simple expressions for the asymptotic bias and variance, by showing that the sandwich smoother is  asymptotically equivalent to a bivariate kernel regression estimator with a product kernel.
As far as we are aware, this is the first central limit theorem for a bivariate spline estimator of any type. Our simulation study shows that the sandwich smoother is orders of magnitude faster to compute than other bivariate spline smoothers, even when the latter are computed using a fast GLAM (Generalized Linear Array Model) algorithm, and
comparable to them  in terms of mean squared integrated errors. We  extend the sandwich smoother to array data of higher dimensions, where a GLAM algorithm improves the computational speed of the sandwich smoother. One important application of  the sandwich smoother is to  estimate  covariance functions in functional data analysis.  In this application, our numerical results show that the sandwich smoother is orders of magnitude faster than local linear regression.  The speed of the sandwich formula is important because functional data sets are becoming quite large.\\

\end{abstract}
\newpage
\baselineskip 22pt

\section{Introduction}\label{sec:introduction}
This paper introduces a fast penalized spline method for
bivariate smoothing. It also gives the first local central limit theorem for a bivariate spline smoother. Suppose there is a regression function $\mu(x,z)$ with $(x,z)\in [0,1]^2$. Initially we assume that  $y_{i,j} = \mu(x_i,z_j)+\epsilon_{i,j}, 1\leq i \leq n_1, 1\leq j\leq n_2$,  where the $\epsilon_{i,j}$'s are independent with $\textrm{E}\epsilon_{i,j} = 0$ and  $\textrm{E}\epsilon_{i,j}^2 =\sigma^2(x_i,z_j)$, and the design points $\{(x_i,z_j)\}_{1\leq i\leq n_1, 1\leq j\leq n_2}$ are deterministic; thus, the total number of data points is $n=n_1n_2$ and the data are on a rectangular grid.  In Section~\ref{sec:irregular} we relax the design assumption to fixed design points not in a regular grid and random design points. With the data on a rectangular grid, they can be organized into an $n_1\times n_2$ matrix $\mathbf{Y}$.  We propose to smooth across the rows and down the columns of  $\mathbf{Y}$ so that the matrix of fitted values $\hat{\mathbf{Y}}$ satisfies
\begin{equation}
\label{HatY}
\hat{\mathbf{Y}} = \mathbf{S}_1\mathbf{Y}\mathbf{S}_2,
\end{equation}
where $\mathbf{S}_1$ ($\mathbf{S}_2$) is the smoother matrix for $x$ ($z$). So fixing one covariate, we smooth along the other covariate and vice versa, although the two smooths are simultaneous as implied by~(\ref{HatY}). Estimator
\eqref{HatY} is similar in form to the sandwich formula for a covariance matrix, which suggested the name
``sandwich smoother" to a referee.  We have adopted this term. 

The tensor product structure of the sandwich smoother allows  fast computations, specifically of the generalized cross validation (GCV) criterion for selecting smoothing parameters; see Section~\ref{sec:algorithm}.
Dierckx (1982) proposed a smoother with the same structure as~(\ref{HatY}), but our asymptotic analysis and the fast implementation for the sandwich smoother are new. For smoothing two-dimensional histograms, Eilers and Goeman (2004) studied a simplified version of the sandwich smoother with special smoother matrices that lead to  non-negative smooth for non-negative data. The fast method for the sandwich smoother can be applied to their method.

For bivariate spline smoothing, there are two well known estimators:  bivariate P-splines (Eilers and Marx, 2003; Marx and Eilers, 2005)
and thin plate splines, e.g., the thin plate regression splines (Wood, 2003). For convenience,  the Eilers-Marx  and Wood estimators will be denoted by E-M  and TPRS, respectively. We use E-M without specification of how the estimator is calculated.

Penalized splines have become popular over the years,  as they use fewer  knots and  in higher dimensions require much less computation than smoothing splines or thin plate splines. See Ruppert {\it et al.}\ (2003) or Wood (2006) for both methodological development and applications.  However, the theoretical study of penalized splines  has been challenging. An asymptotic study of univariate penalized splines was achieved only recently  (Opsomer and Hall, 2005; Li and Ruppert, 2008; Claeskens {\it et al.}, 2009; Kauermann {\it et al.}, 2009; Wang {\it et al.}, 2011). The asymptotic convergence rate of smoothing splines, on the other hand, has been well established; see  Gu (2002) for a comprehensive list of references.

The theoretical study of penalized splines in higher dimension is  more challenging.
To the best of our knowledge, the  literature does not contain central limit theorems or
explicit expressions for the asymptotic mean and covariance matrix of $\hat \mu(x,z)$ for  bivariate spline estimators of any kind.
The sandwich smoother has a tensor product structure that simplifies asymptotic analysis, and we show that the sandwich smoother is asymptotically equivalent to
a kernel estimator with a product kernel.  Using this result, we obtain a central limit theorem  for the sandwich smoother and simple expressions
for the asymptotic bias and variance.  

For smoothing of array data, the generalized linear array model (GLAM) by Currie {\it et al.}\,(2006) gives a low storage, high speed algorithm by making use of the array structures of the model matrix and the data. The E-M estimator can be implemented with a GLAM algorithm (denoted by E-M/GLAM). The sandwich smoother can also be extended to array data of arbitrary dimensions where a GLAM algorithm  can improve the speed of the sandwich smoother; see Section~\ref{sec:multi}. Because of the fast methods in Sections~\ref{sec:algorithm} and \ref{sec:multi:algorithm} for computing the GCV criterion, a GLAM algorithn is much faster when used to calculate the sandwich smoother than when used to calculate the E-M estimator.   In Table~\ref{time} in Section~\ref{sec:speed}, we see that the sandwich smoother is many orders of magnitude faster than the E-M/GLAM estimator over a wide range of sample sizes and numbers of knots.

The remainder of this paper is organized as follows. In Section~\ref{sec:bivariate}, we give details about the sandwich smoother.  In Section~\ref{sec:asymptotics}, we establish an asymptotic theory of the sandwich smoother by showing that it is asymptotically equivalent to a bivariate kernel
estimator with a product kernel. In Section~\ref{sec:irregular}, we consider irregularly spaced data.
 In Section~\ref{sec:simulations}, we report a
simulation study.
In Section~\ref{sec:covariance}, we compare the sandwich smoother with a local linear smoother for estimating covariance functions of functional data.
We find that the sandwich smoother is many orders of magnitude faster than the local linear smoother and they have similar mean integrated squared errors (MISEs).
In Section~\ref{sec:multi}, we extend the sandwich smoother to array data of dimension greater than two.

 \section{The sandwich smoother}\label{sec:bivariate}
 Let \vec\, be  the operation that stacks the columns of a matrix into a vector. Define $\mathbf{y}=\vec(\mathbf{Y})$ and $\vec(\hat{\mathbf{Y}})=\hat{\mathbf{y}}$. Applying  a well-known identity of the tensor product (Seber 2007, pp.\ 240) to~(\ref{HatY}) gives
\begin{equation}
\label{Haty}
\hat{\mathbf{y}}=(\mathbf{S}_2\otimes \mathbf{S}_1)\mathbf{y}.
\end{equation}
Identity~(\ref{Haty}) shows that the overall smoother matrix is a tensor product of two univariate smoother matrices. Because of this factorization of the smoother matrix, we say our model has a tensor product structure. We shall use  P-splines (Eilers and Marx, 1996) to construct univariate smoother matrices, i.e.,
\begin{equation}
\label{S}
 \mathbf{S}_i = \mathbf{B}_i(\mathbf{B}_i^T\mathbf{B}_i+\lambda_i \mathbf{D}_i^T\mathbf{D}_i)^{-1}\mathbf{B}_i^T, i=1,2,
 \end{equation}
  where $\mathbf{B}_1$ and $\mathbf{B}_2$ are the model matrix for $x$ and $z$ using B-spline basis (defined later), and $\mathbf{D}_1$ and $\mathbf{D}_2$ are differencing matrices of difference orders $m_1$ and $m_2$, respectively.  Then the overall smoother matrix can be written out  using identities of the tensor product (Seber 2007, pp.\ 235-239),
\begin{equation}
\label{hat2}
\begin{split}
\mathbf{S}_2\otimes\mathbf{S}_1&=\left\{\mathbf{B}_2(\mathbf{B}_2^T\mathbf{B}_2+\lambda_2 \mathbf{D}_2^T\mathbf{D}_2)^{-1}\mathbf{B}_2^T\right\}\otimes \left\{\mathbf{B}_1(\mathbf{B}_1^T\mathbf{B}_1+\lambda_1 \mathbf{D}_1^T\mathbf{D}_1)^{-1}\mathbf{B}_1^T\right\}\\
&=(\mathbf{B}_2\otimes\mathbf{B}_1)\{\mathbf{B}_2^T\mathbf{B}_2\otimes\mathbf{B}_1^T\mathbf{B}_1+\lambda_1\mathbf{B}_2^T\mathbf{B}_2\otimes \mathbf{D}_1^T\mathbf{D}_1\\
&\quad\,+\lambda_2 \mathbf{D}_2^T\mathbf{D}_2\otimes \mathbf{B}_1^T\mathbf{B}_1
+\lambda_1\lambda_2 \mathbf{D}_2^T\mathbf{D}_2\otimes \mathbf{D}_1^T\mathbf{D}_1\}^{-1}(\mathbf{B}_2\otimes\mathbf{B}_1)^T.
\end{split}
\end{equation}

The inverse matrix in the second equality of~(\ref{hat2}) shows that our model uses tensor-product splines (defined later) with penalty
\begin{equation}
\label{P}
\mathbf{P} = \lambda_1\mathbf{B}_2^T\mathbf{B}_2\otimes \mathbf{D}_1^T\mathbf{D}_1+\lambda_2 \mathbf{D}_2^T\mathbf{D}_2\otimes \mathbf{B}_1^T\mathbf{B}_1
+\lambda_1\lambda_2 \mathbf{D}_2^T\mathbf{D}_2\otimes \mathbf{D}_1^T\mathbf{D}_1
\end{equation}
on the coefficients matrix. The tensor-product splines of two variables (Dierckx 1995, ch. 2) is defined by
\[
\sum_{1\leq \kappa\leq c_1,1\leq \ell\leq c_2} \theta_{\kappa,\ell}B^1_{\kappa}(x)B^2_{\ell}(z),
\]
where $B^1_{\kappa}$ and $B^2_{\ell}$ are B-spline basis functions for $x$ and $z$, respectively, $c_1$ and $c_2$ are the numbers of basis functions for the univariate splines, and $\boldsymbol{\Theta}=(\theta_{\kappa,\ell})_{1\leq \kappa\leq c_1,1\leq \ell\leq c_2}$ is the coefficients matrix. We use B-splines of degrees $p_1$ ($p_2$) for $x$ ($z$), and use $K_1-1$ ($K_2-1$) equidistant interior knots.  Then $c_1=K_1+p_1$, $c_2=K_2+p_2$. It
 follows that the model is
 \begin{equation}
 \label{model}
\mathbf{Y}=\mathbf{B}_1\mathbf{\Theta}\mathbf{B}_2^T+\boldsymbol{\epsilon},
 \end{equation}
 where $\mathbf{B}_1=\{B^1_{\kappa}(x_r)\}_{1\leq r\leq  n_1, 1\leq \kappa \leq c_1}$, $\mathbf{B}_2=\{B^2_{\ell}(z_s)\}_{1\leq s\leq  n_2, 1\leq \ell \leq c_2}$, and $\boldsymbol{\epsilon}$ is an $n_1\times n_2$ matrix with $(i,j)th$ entry $\epsilon_{i,j}$. Let $\boldsymbol{\theta}=$\vec($\boldsymbol{\Theta}$). Then an estimate of $\boldsymbol{\theta}$ is given by minimizing
$
\| \mathbf{Y}-\mathbf{B}_1\hat{\boldsymbol{\Theta}}\mathbf{B}_2^T\|_F^2+\hat{\boldsymbol{\theta}}^T\mathbf{P}\hat{\boldsymbol{\theta}},
$
where the norm is the Frobenius norm and $\mathbf{P}$ is defined in~(\ref{P}). It follows that the estimate of the coefficient matrix $\hat{\boldsymbol{\Theta}}$ satisfies
$\mathbf{\Lambda}_1\hat{\boldsymbol{\Theta}}\mathbf{\Lambda}_2= \mathbf{B}_1^T\mathbf{Y}\mathbf{B}_2,
$
where  for $i=1, 2$,  $\mathbf{\Lambda}_i= \mathbf{B}_i^T\mathbf{B}_i+\lambda_i
\mathbf{D}_i^T\mathbf{D}_i$, or equivalently, $\hat{\boldsymbol{\theta}}$ satisfies
\begin{equation}
\label{eqnobj}
\left(\mathbf{\Lambda}_2\otimes \mathbf{\Lambda}_1\right)\hat{\boldsymbol{\theta}}=  (\mathbf{B}_2\otimes \mathbf{B}_1)^T\mathbf{y}.
\end{equation}
Then our penalized estimate is
\begin{equation}
\label{est}
\hat{\mu}(x,z)= \sum_{1\leq \kappa\leq c_1,1\leq \ell\leq c_2} \hat{\theta}_{\kappa,\ell}B^1_{\kappa}(x)B^2_{\ell}(z).
\end{equation}
With~(\ref{eqnobj}), it is straightforward to show that $\hat{\mathbf{y}}=(\mathbf{B}_2\otimes \mathbf{B}_1)\hat{\boldsymbol{\theta}}$ satisfies~(\ref{HatY}), which confirms that the proposed method uses tensor-product splines with a particular penalty.

\subsection{Comparison with the E-M estimator}\label{sec:comparison}
The only difference between the sandwich smoother and the E-M estimator (Marx and Eilers, 2003; Eilers and Marx, 2006) is the penalty.  Let $\mathbf{P}_{\text{E-M}}$ denote the penalty matrix for the E-M estimator,  then $\mathbf{P}_{\text{E-M}}= \lambda_1\mathbf{I}_{c_2}\otimes \mathbf{D}_1^T\mathbf{D}_1
+\lambda_2
\mathbf{D}_2^T\mathbf{D}_2\otimes \mathbf{I}_{c_1}$. The first and second  penalty terms in bivariate P-splines penalize the columns and rows of $\boldsymbol{\Theta}$, respectively, and are thus called column and row penalties.   It can be shown that the first penalty term in~(\ref{P}),   $\mathbf{B}_2^T\mathbf{B}_2\otimes \mathbf{D}_1^T\mathbf{D}_1$, like $\mathbf{I}_{c_2}\otimes \mathbf{D}_1^T\mathbf{D}_1$, is  a ``column" penalty, but it penalizes the columns of $\boldsymbol{\Theta}\mathbf{B}_2^T$ instead of the columns of $\boldsymbol{\Theta}$. We call this a modified column penalty. The implication of this modified column penalty can be seen from a closer look at model~(\ref{model}). By regarding  (\ref{model}) as a model with B-spline base $\mathbf{B}_1$ and coefficients $\boldsymbol{\Theta}\mathbf{B}_2^T$, (\ref{model}) becomes a varying-coefficients model (Hastie and Tibshirani, 1993) in $x$ with coefficients depending on $z$. So we can interpret the modified column penalty as a penalty for the  univariate P-spline smoothing along the $x$-axis. Similarly, the penalty term $ \mathbf{D}_2^T\mathbf{D}_2\otimes \mathbf{B}_1^T\mathbf{B}_1$ for the sandwich smoother penalizes the rows of $\mathbf{B}_1\boldsymbol{\Theta}$ and  can be interpreted as the penalty for the univariate P-spline smoothing along the $z$-axis. The third penalty in~(\ref{hat2}) corresponds to the interaction of the two univariate smoothing.

\subsection{A fast implementation}\label{sec:algorithm}
We derive a fast implementation for the sandwich smoother by showing how the smoothing parameters can be selected  via a fast computation of \mbox{GCV}. GCV requires the computation of  $\|\hat{\mathbf{Y}}-\mathbf{Y}\|_F^2$ and  the trace of the overall smoother matrix.  We need some initial computations. First,  we need the singular valued decompositions
\begin{equation}
(\mathbf{B}_i^T\mathbf{B}_i)^{-1/2}\mathbf{D}_i^T\mathbf{D}_i(\mathbf{B}_i^T\mathbf{B}_i)^{-1/2} = \mathbf{U}_i \textrm{diag}(\mathbf{s}_i)\mathbf{U}_i^T,\quad\text{for}\,\, i = 1, 2,\label{eq:DR01}
\end{equation}
where $\mathbf{U}_i $ is the matrix of eigenvectors and $\mathbf{s}_i$ is the vector of eigenvalues. For $i =1 ,2$,  let $
\mathbf{A}_i = \mathbf{B}_i(\mathbf{B}_i^T\mathbf{B}_i)^{-1/2}\mathbf{U}_i$,
then $\mathbf{A}_i^T\mathbf{A}_i = \mathbf{I}_{c_i}$ and $\mathbf{A}_i \mathbf{A}_i^T = \mathbf{B}_i(\mathbf{B}_i^T\mathbf{B}_i)^{-1}\mathbf{B}_i^T$.  It follows that  for $i=1, 2$,
$
\mathbf{S}_i = \mathbf{A}_i\boldsymbol{\Sigma}_i \mathbf{A}_i^T
$
with $\boldsymbol{\Sigma}_i = \left\{\mathbf{I}_{c_i}+\lambda_i \textrm{diag}(\mathbf{s}_i)\right\}^{-1}$.

 We first  compute $\|\hat{\mathbf{Y}}-\mathbf{Y}\|_F^2$.
 Substituting $ \mathbf{A}_i\boldsymbol{\Sigma}_i \mathbf{A}_i^T$ for $\mathbf{S}_i$ in  equation~(\ref{HatY}) we obtain
 \begin{equation*}
 \hat{\mathbf{Y}} = \mathbf{A}_1\left\{\boldsymbol{\Sigma_1} \left(\mathbf{A}_1^T\mathbf{Y}\mathbf{A}_2\right)\boldsymbol{\Sigma_2}\right\}\mathbf{A}_2^T =  \mathbf{A}_1\left(\boldsymbol{\Sigma_1} \tilde{\mathbf{Y}}\boldsymbol{\Sigma_2}\right)\mathbf{A}_2^T,
 \end{equation*}
 where $\tilde{\mathbf{Y}}= \mathbf{A}_1^T\mathbf{Y}\mathbf{A}_2$.  Let $\tilde{\mathbf{y}} = \vec(\tilde{\mathbf{Y}})$, then
 \begin{equation}
 \label{gcv01}
 \hat{\mathbf{y}} = (\mathbf{A}_2\otimes\mathbf{A}_1)(\boldsymbol{\Sigma}_2\otimes\boldsymbol{\Sigma}_1)\tilde{\mathbf{y}}.
 \end{equation}
 We shall use the following operations on vectors: let $\mathbf{a}$ be a vector containing  only positive elements,  $\mathbf{a}^{1/2}$ denotes the element-wise squared root of $\mathbf{a}$ and $1/\mathbf{a}$ denotes the element-wise inverses of $\mathbf{a}$. 
We can derive that
  \begin{equation}
 \label{gcv0}
   \|\hat{\mathbf{Y}}-\mathbf{Y}\|_F^2 =\left\{ \tilde{\mathbf{y}}^T\left (\tilde{\mathbf{s}}_2\otimes\tilde{\mathbf{s}}_1\right)\right\}^2 - 2\left\{ \tilde{\mathbf{y}}^T \left(\tilde{\mathbf{s}}_2^{1/2}\otimes\tilde{\mathbf{s}}_1^{1/2}\right)\right\}^2 +\mathbf{y}^T\mathbf{y},
    \end{equation}
 where $\tilde{\mathbf{s}}_i = 1/ (\mathbf{1}_{c_i} + \lambda_i \mathbf{s}_i)$ for $i=1, 2$ and $\mathbf{1}_{c_i}$ is a vector of $1$'s  with length $c_i$. See Appendix~\ref{sec:derivation} for the derivation of~(\ref{gcv0}). The right hand of~(\ref{gcv0}) shows that for each pair of smoothing parameters the calculation of $ \|\hat{\mathbf{Y}}-\mathbf{Y}\|_F^2$ is just two inner product of vectors of length $c_2c_1$ and the term $\mathbf{y}^T\mathbf{y}$ just needs one calculation for all smoothing parameters.

 Next,  the trace of the overall smoother matrix  can be computed by  first using another identity of the tensor product (Seber 2007, pp.\ 235)
\begin{equation}
\label{trace}
\tr (\mathbf{S}_2\otimes\mathbf{S}_1) = \tr(\mathbf{S}_2)\cdot\tr(\mathbf{S}_1),
\end{equation}
and then using a trace identity ${\rm tr}(AB) = {\rm tr}(BA)$ (if the dimensions are compatible) (Seber, 2007, pp.\ 55) and as well as the fact that $\mathbf{A}_i^T\mathbf{A}_i=\mathbf{I}_{c_i}$,
\begin{equation}
\label{trace2}
\tr(\mathbf{S}_i) = \sum_{\kappa=1}^{c_i}\frac{1}{1+\lambda_i s_{i,\kappa}},
\end{equation}
where $s_{i,\kappa}$ is the $\kappa$th element of $\mathbf{s}_i$.

To summarize, by equations (\ref{gcv0}),  (\ref{trace}) and (\ref{trace2}) we obtain a fast implementation for computing GCV that enables us to select the smoothing parameters efficiently. Because of the fast implementation, the sandwich smoother can be much faster than the E-M/GLAM algorithm; see Section \ref{sec:speed} for an empirical comparison. For the E-M/GLAM estimator, the inverse of a matrix of dimension $c_1c_2\times c_1c_2$ is required for every pair of $(\lambda_1,\lambda_2)$, while for the sandwich smoother, except in the initial computations in (\ref{eq:DR01}), no matrix inversion is required.


\def\m3{m_3}
\section{Asymptotic theory} \label{sec:asymptotics}
In this section, we derive the asymptotic distribution of  the sandwich smoother and show that
it is asymptotically equivalent to a bivariate kernel regression estimator with a product kernel. Moreover, we show that when the  two orders of difference penalties are the same, the sandwich smoother  has the optimal rate of convergence.

We shall use the equivalent kernel method first used for studying smoothing splines (Silverman, 1984) and also useful in studying the asymptotics of P-splines (Li and Ruppert, 2008; Wang {\it et al}., 2011).   A nonparametric point estimate is usually a weighted average of all data points, with the weights depending on the point and the method being used. The equivalent kernel method  shows that the weights are asymptotically the weights from  a kernel regression estimator for some kernel function (the equivalent kernel) and some bandwidth (the equivalent bandwidth). First, we define a univariate kernel
function
\begin{equation} H_{m}(x)=\sum_{\nu=1}^m
\frac{\psi_{\nu} }{2m}\exp\{-\psi_\nu
|x|\},\label{kernel}
\end{equation} where $m$ is a positive integer and
the $\psi_{\nu} $'s are the $m$ complex roots of
$x^{2m}+(-1)^m =0$ that have positive real parts. Here $H_m$ is the equivalent  kernel  for univariate penalized splines (Wang {\it et al.}, 2011). By Lemma~\ref{H_m} in Appendix~\ref{sec:proof}, $H_m$ is of order $2m$. Note that the order of a kernel determines the convergence rate of the kernel estimator. See Wand and Jones (1995) for more details.  A bivariate kernel regression estimator  with the product kernel $H_{m_1}(x)H_{m_2}(z)$ is of the form $(n h_{n,1}h_{n,2})^{-1} \sum_{i,j} y_{i,j} H_{m_1}\left\{h_{n,1}^{-1}(x-x_i)\right\}H_{m_2}\left\{h_{n,2}^{-1}(z-z_j)\right\}$, where $h_{n,1}$ and $h_{n,2}$ are the bandwidths. Under appropriate assumptions, the sandwich smoother  is asymptotically equivalent to the above kernel estimator (Proposition~\ref{prop1}). Because the asymptotic theory of a kernel regression estimator is well established (Wand and Jones, 1995),  an asymptotic theory can be similarly established  for the sandwich smoother. For notational convenience, $a\sim b$ implies $a/b$ converges to 1.
\begin{prop}\label{prop1}
Assume the following conditions are satisfied.
\begin{enumerate}
\item There exists a constant $\delta>0$ such that
$\sup_{i,j}\textrm{E}\left(|y_{i,j}|^{2+\delta}\right)<\infty$.  \item The regression function
$\mu(x,z)$ has continuous $2m$th order derivatives where
$m=\max(m_1,m_2)$.
\item The variance function $\sigma^2(x,z)$ is continuous.
 \item The
covariates satisfy $(x_i,z_j) = ((i-1/2)/n_1, (j-1/2)/n_2)$.
\item $n_1\sim c n_2$ where $c$ is a constant.
\end{enumerate}
Let $h_{n,1}=K_1^{-1}(\lambda_1 K_1 n_1^{-1})^{1/(2m_1)}$, $h_{n,2}=K_2^{-1}(\lambda_2 K_2 n_2^{-1})^{1/(2m_2)}$ and $h_n = h_{n,1}h_{n,2}$. Assume  $h_{n,1}=O(n^{-\nu_1})$ and $h_{n,2}=O(n^{-\nu_2})$ for some constants $0<\nu_1,\nu_2<1$. Assume also $(K_1h_{n,1}^2)^{-1} = o(1)$ and $(K_2h_{n,2}^2)^{-1} = o(1)$.  Let $\hat{\mu}(x,z)$ be the sandwich smoother using
$m_1$th ($m_2$th) order difference penalty and $p_1\geq 1$
($p_2\geq 1$) degree B-splines on the $x$-axis ($z$-axis) with
equally spaced knots. Fix $(x,z)\in (0,1)\times (0,1)$.  \\Let $ \mu^{\ast}(x,z) = (n h_{n})^{-1}\sum_{i,j} y_{i,j} H_{m_1}\left\{h_{n,1}^{-1}(x-x_i)\right\}H_{m_2}\left\{h_{n,2}^{-1}(z-z_j)\right\}$. Then
\begin{align*}
\textrm{\textnormal{E}}\left\{\hat\mu(x,z)-\mu^{\ast}(x,z)\right\} &= O\left[\max\{(K_1h_{n,1})^{-2}, (K_2h_{n,2})^{-2}\}\right],\\
\textrm{\textnormal{var}}\{\hat\mu(x,z)-\mu^{\ast}(x,z)\} &= o\{(nh_n)^{-1}\}.
\end{align*}

\end{prop}

All proofs are given in Appendix~\ref{sec:proof}.
\begin{thm}\label{thm1}
Use the same notation in Proposition~\ref{prop1} and assume all conditions and assumptions in Proposition~\ref{prop1} are satisfied.  To simplify notation, let $\m3 = 4m_1 m_2 + m_1 + m_2$. Furthermore, assume that $K_1 \sim  C_1 n^{\tau_1}, K_2 \sim C_2 n^{\tau_2}$
with $\tau_1>{(m_1+1)m_2}/{\m3},$ $\tau_2>{m_1(m_2+1)}/
{\m3}$, $h_{n,1}\sim
h_1n^{-{m_2}/\m3}, h_{n,2}\sim
h_2n^{-{m_1}/m3}$ for positive constants
$C_1,C_2$ and $ h_1, h_2$. Then, for any $(x,z)\in (0,1)\times (0,1)$,
we have that
\begin{equation}
\label{asymptotics}
n^{(2m_1m_2)/\m3}\left\{\hat{\mu}(x,z)-\mu(x,z)\right\}\Rightarrow N\left\{\tilde{\mu}(x,z),V(x,z) \right\}
\end{equation}
in distribution as $n_1\rightarrow\infty, n_2\rightarrow\infty$,  where
\begin{eqnarray}
&&\tilde{\mu}(x,z)  = (-1)^{m_1+1}h_1^{2m_1}\frac{\partial^{2m_1}}{\partial x^{2m_1}}\mu(x,z)+(-1)^{m_2+1}h_2^{2m_2}\frac{\partial^{2m_2}}{\partial z^{2m_2}}\mu(x,z),\label{tilde_mu}\\
&&V(x,z)  =  \sigma^2(x,z)\int H_{m_1}^2(u)\mathrm{d}u \int H_{m_2}^2(v)\mathrm{d}v \label{V(x,z)}.
\end{eqnarray}
\end{thm}

\begin{rem}
The case $m_1=m_2=m$ is important. The convergence rate of the estimator becomes $n^{-m/(2m+1)}$.  Stone (1980) obtained the optimal rates of convergence for nonparametric estimators.  For a bivariate smooth function $\mu(x,z)$  with continuous $2m$th  derivatives, the corresponding optimal rate of convergence  for estimating $\mu(x,z)$ at any inner point of the unit square is $n^{-m/(2m+1)}$. Hence when $m_1=m_2=m$, the sandwich smoother achieves the optimal rate of convergence. Note that the bivariate kernel estimator with the product kernel $H_m(x)H_m(z)$ also has a convergence rate of $n^{-m/(2m+1)}$.
\end{rem}
\begin{rem}
 For the univariate case, the convergence rate of P-splines  with an $m$th order difference penalty is $n^{-2m/(4m+1)}$ (see Wang {\it et al}., 2011). So the rate of convergence for the bivariate case is slower  which shows the effect of ``curse of dimensionality".
  \end{rem}

\begin{rem}
\label{rem3}
Theorem~\ref{thm1} shows that, provided it is fast enough, the divergence rate of the number of knots does not affect the asymptotic distribution. For practical usage, we recommend $K_1 = \min\{n_1/2, 35\}$ and $K_2 = \min\{n_2/2, 35\}$,  so that every bin has at least 4 data points. Note that for univariate P-splines, a number of $\min\{n/4, 35\}$ knots was recommended by Ruppert (2002).
\end{rem}

\section{Irregularly spaced data}\label{sec:irregular}
Suppose the design points are random and  we use the model $y_i = \mu(x_i,z_i) + \epsilon_i, i=1,\dots, n$, that is $y_i$, $x_i$,
and $z_i$ now have only a single index rather than $i,j$ as before.
Assume the design points $\{(x_1, z_1),\dots, (x_n, z_n)\}$ are independent and sampled from a distribution $F(x,z)$ in $[0,1]^2$.  The \mbox{sandwich smoother}  can
not be directly applied to irregularly spaced data. A solution to this problem is to bin the data first.  We partition $[0, 1]^2$ into an $I_1\times I_2$ grid of
equal-size rectangular bins, and let  $\tilde{y}_{\kappa,\ell}$ be the mean of all $y_i$ such that $(x_i,z_i)$ is in the $(\kappa,\ell)$th bin. If there are no data in the $(\kappa,\ell)$th bin, $\tilde{y}_{\kappa,\ell}$ is defined arbitrarily, e.g.,
 by a nearest neighbor estimator (see below).  Assuming $\tilde{y}_{\kappa, \ell}$ is a data point  at  $(\tilde x_{\kappa}, \tilde z_{\ell})$, the center of the $(\kappa,\ell)$th bin,  we apply the sandwich smoother to the grid data $\tilde{\mathbf{Y}}= (\tilde{y}_{\kappa,\ell})_{1\leq \kappa\leq I_1,1\leq \ell\leq I_2}$ to get
$$
\hat{\boldsymbol{\theta}}^{\ast}= \left(\boldsymbol{\Lambda}_2^{-1}\otimes\boldsymbol{\Lambda}_1^{-1}\right)\left (\mathbf{B}_2\otimes \mathbf{B}_1\right)^T\mathbf{\tilde{\mathbf{y}}},
$$
where $\tilde{\mathbf{y}}=$ \vec$(\tilde{\mathbf{Y}})$. Then our penalized estimate is defined as
\begin{equation*}
\hat\mu(x,z)=\sum_{\kappa=1}^{c_1}\sum_{\ell=1}^{c_2}\hat
\theta^{\ast}_{k,\ell}B_{\kappa}^1(x)B_{\ell}^2(z).
\end{equation*}
\subsection{Practical implementation}
For the above estimation procedure  to work with the fast implementation in
Section~\ref{sec:algorithm}, we need to handle the problem when there are no data in some bins due to sampling variation. If there are no data in the $(\kappa, \ell)$th bin, one solution is to define $\tilde{y}_{\kappa,\ell}$ to be the mean of values in the neighboring bins. Doing this has no effect on asymptotics, since bins  will eventually have data. For small samples, filling in empty cells this way allows the sandwich smoother to be calculated, but one might flag the estimates in the vicinity of empty bins as non-reliable.

Another solution is to  use an algorithm which iterates between the data and the smoothing parameters as follows. Initially, we let
$\tilde{y}_{\kappa,\ell}=0$ if the $(\kappa,\ell)$th bin has no data point.
Another possibility is to let $\tilde{y}_{\kappa,\ell}$ be, for some $M>0$, the average of the $M$  values of $y$ with
$(x,z)$ coordinates located closest to the center
of the $(\kappa,\ell)$th bin. To determine the smoothing
parameters $(\lambda_1,\lambda_2)$ that minimize GCV, we only calculate the sums of squared errors  for the bins with data and ignore the bins with no data. This gives us an initial pair of smoothing parameters. Then for the bins
with no data,  we replace the $\tilde{y}_{\kappa,\ell}$'s by the estimated value with this pair of smoothing parameters. Now with the updated data, we could obtain another pair of smoothing parameters. We repeat the above procedure until reaching some convergence.


\subsection{Asymptotic theory}
As before, we divide the unit interval into an $I_1\times I_2$ grid and let $I=I_1I_2$ be the number of bins. 
\begin{thm}\label{thm2}
Assume the following conditions are satisfied.
\begin{enumerate}
\item There exists a constant $\delta>0$ such that
$\sup_{i}\textrm{E}\left(|y_i|^{2+\delta}\right)<\infty$.
\item The regression function
$\mu(x,z)$ has continuous  $2m$th order derivatives  where
$m=\max(m_1,m_2)$.
\item The design points $\{(x_i,z_i)\}_{i=1}^n$ are independent and sampled from a distribution $F(x,z)$ with a density function $f(x,z)$ and $f(x,z)$ is positive over $[0,1]^2$ and has continuous first derivatives.
\item Conditional on $\{(x_i,z_i)\}_{i=1}^n$, the random errors $\epsilon_i, 1\leq i\leq n $,  are independent with mean 0 and conditional variance $\sigma^2(x_i, z_i)$.
 \item The variance function $\sigma^2(x,z)$ is twice continuously differentiable.
 \item $I \sim c_I n^{\tau}$ and $I_1 \sim c_0 I_2$ for some constants $c_I, c_0$ and $\tau>(4m_1m_2)/(4m_1m_2+m_1+m_2)$.
\end{enumerate}
Fix $(x,z)\in (0,1)^2$. Then with the same notation and assumptions as in Theorem~\ref{thm1},   we have that
\[
n^{(2m_1m_2)/\m3}\left\{\hat{\mu}(x,z)-\mu(x,z))\right\}\Rightarrow N\left\{\tilde{\mu}(x,z),V(x,z)/f(x,z) \right\}
\]
in distribution as $n \rightarrow\infty$ where $\tilde{\mu}(x,z)$ is defined in~(\ref{tilde_mu}) and $V(x,z)$ is defined in~(\ref{V(x,z)}).
\end{thm}
\begin{rem}
We assume random design points in Theorem~\ref{thm2}. For the fixed design points, the result in Theorem~\ref{thm2} still holds if we replace condition (c) with the following:  $\sup_{\kappa,\ell}\left|n_{\kappa,\ell}/(nI^{-1}) - f(\tilde{x}_{\kappa},\tilde{z}_{\ell}) \right| = o(1)$ where $n_{\kappa,\ell}$ is the number of data points in the $(\kappa,\ell)$th bin and $f(x,z)$ is a continuous and positive function.
\end{rem}

\section{A simulation study}\label{sec:simulations}
This section compares  the sandwich smoother, Eilers and Marx's P-splines implemented with a GLAM algorithm (E-M/GLAM) and Wood's thin-plate regression splines
(TPRS) in terms of mean integrated square errors (MISEs) and computation speed. Section~\ref{sec:regression} shows that  MISEs of the sandwich smoother and E-M/GLAM are roughly comparable and smaller than those of \mbox{TPRS}, while Section~\ref{sec:speed} illustrates the computational advantage of the sandwich smoother over the other smoothers.

\subsection{Regression function estimation}\label{sec:regression}
\begin{figure}
\begin{center}
\subfigure{\includegraphics[ height=2.5in, angle=270]{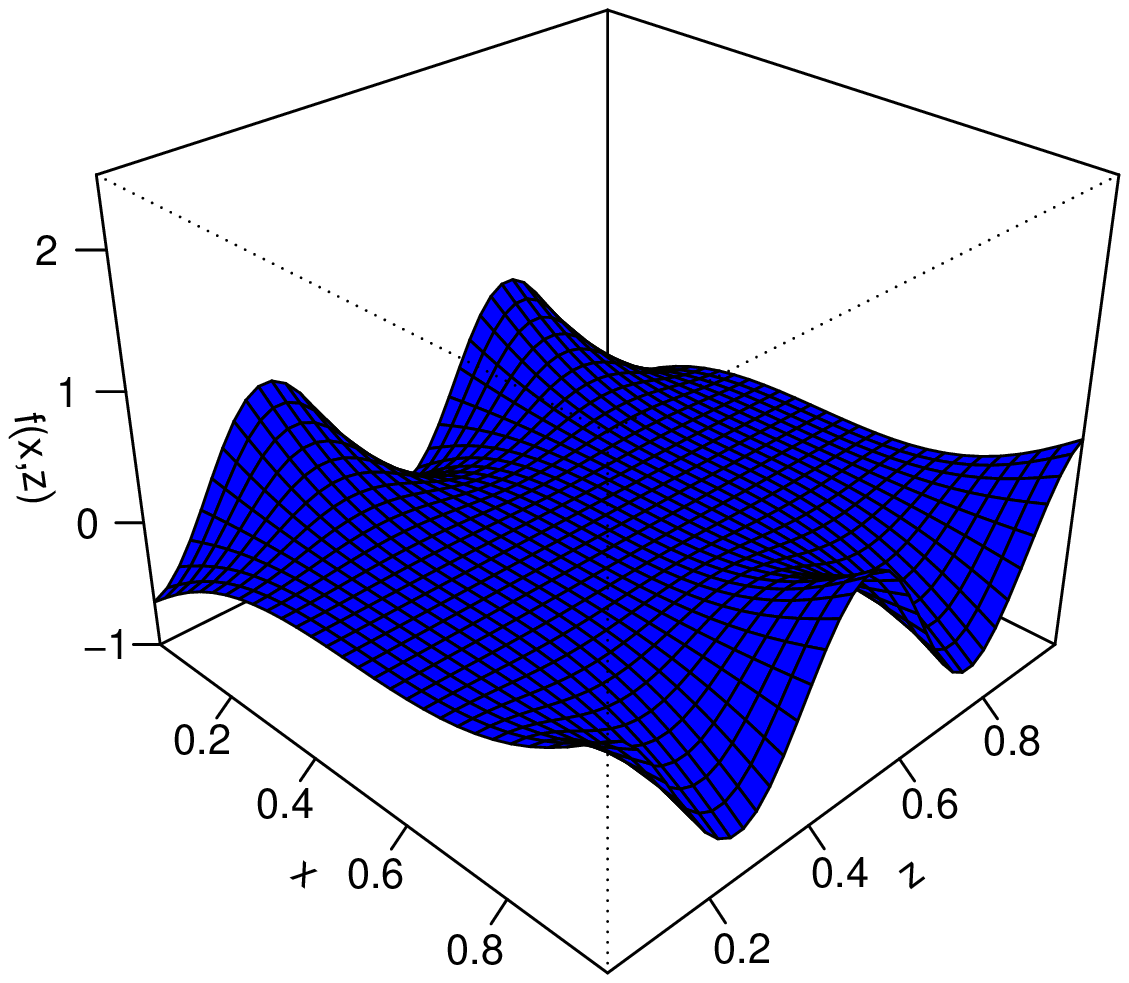}}
\subfigure{\includegraphics[ height=2.5in,  angle=270]{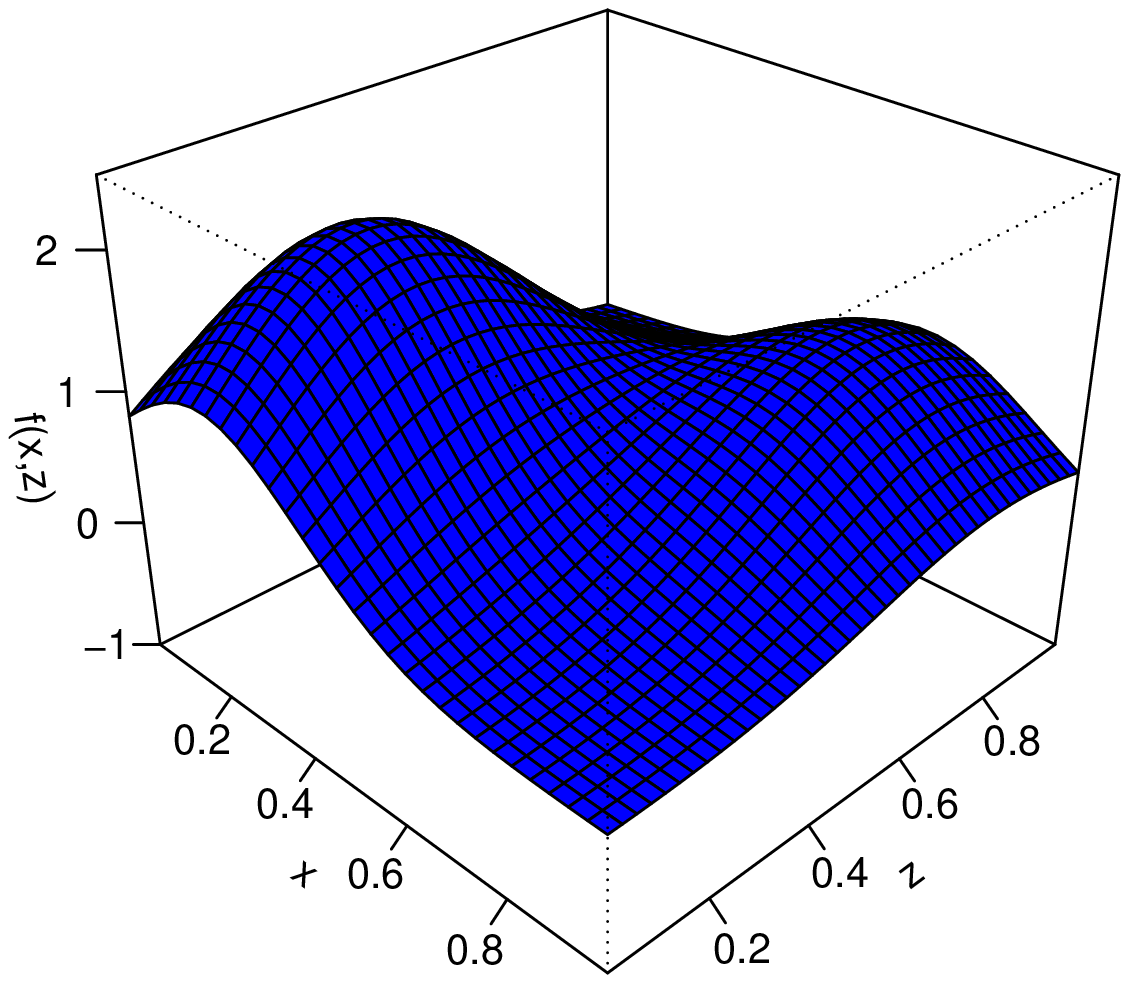}}
\end{center}
\caption{\label{fig0}Surfaces of $f_1$ and $f_2$. The left surface is for $f_1$ and the right one is for $f_2$.}
\end{figure}

Two test functions were used in the simulation study: $f_1(x,z)=\sin\{2\pi (x-.5)^3\}\cos(4\pi z)$ and
\begin{align*}
f_2(x,z) &= \frac{0.75}{\pi \sigma_x\sigma_z} \exp\{-(x-0.2)^2/\sigma_x^2-(z-0.3)^2/\sigma_z^2\}\\
            &\quad+
                  \frac{0.45}{\pi \sigma_x\sigma_z} \exp\{-(x-0.7)^2/\sigma_x^2-(z-0.8)^2/\sigma_z^2\},
\end{align*}
where $\sigma_x = 0.3, \sigma_z =0.4$. Note that $f_2$ was used in Wood (2003). The two true surfaces are shown in Figure~\ref{fig0}.

\begin{figure}
\begin{center}
\subfigure[$f_1,\sigma=0.1$]{\includegraphics[ height=2in, angle=270]{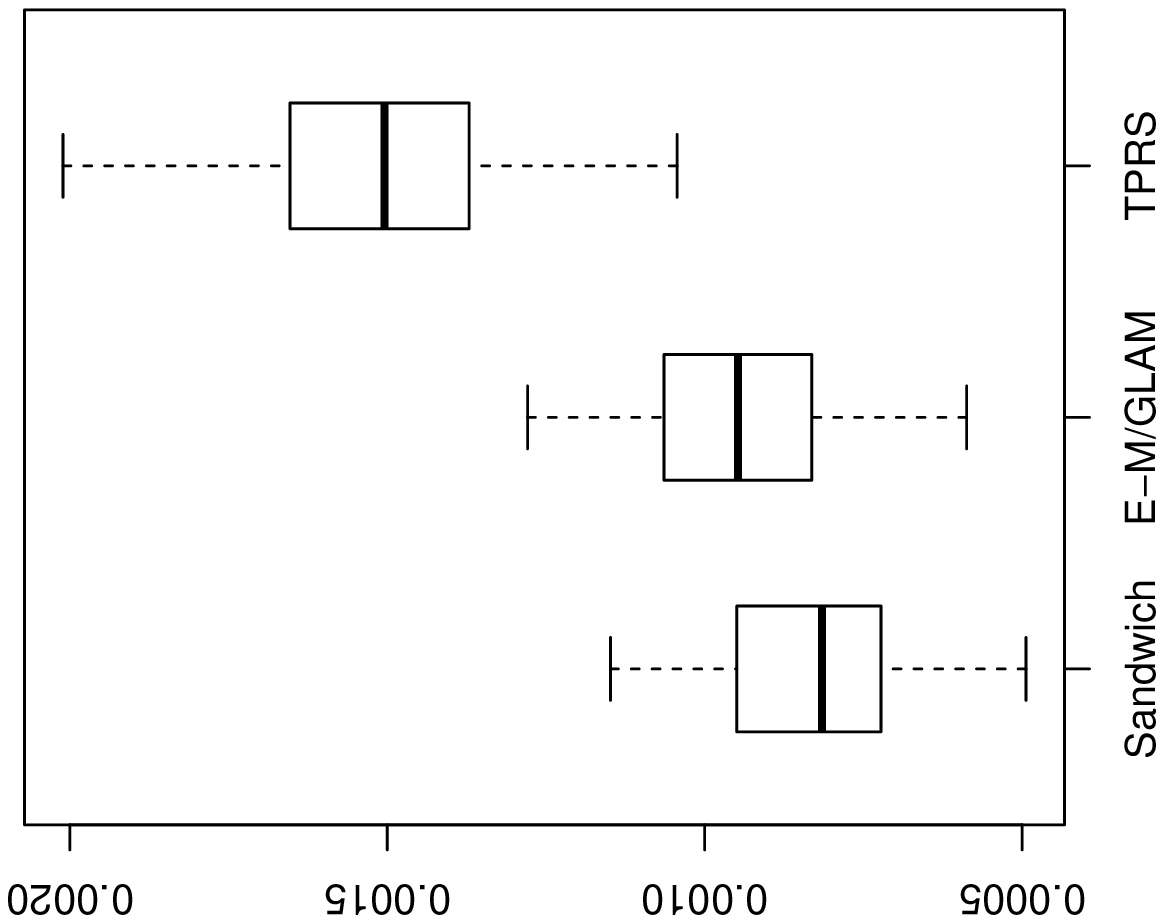}}
\subfigure[$f_1,\sigma=0.5$]{\includegraphics[height=2in,  angle=270]{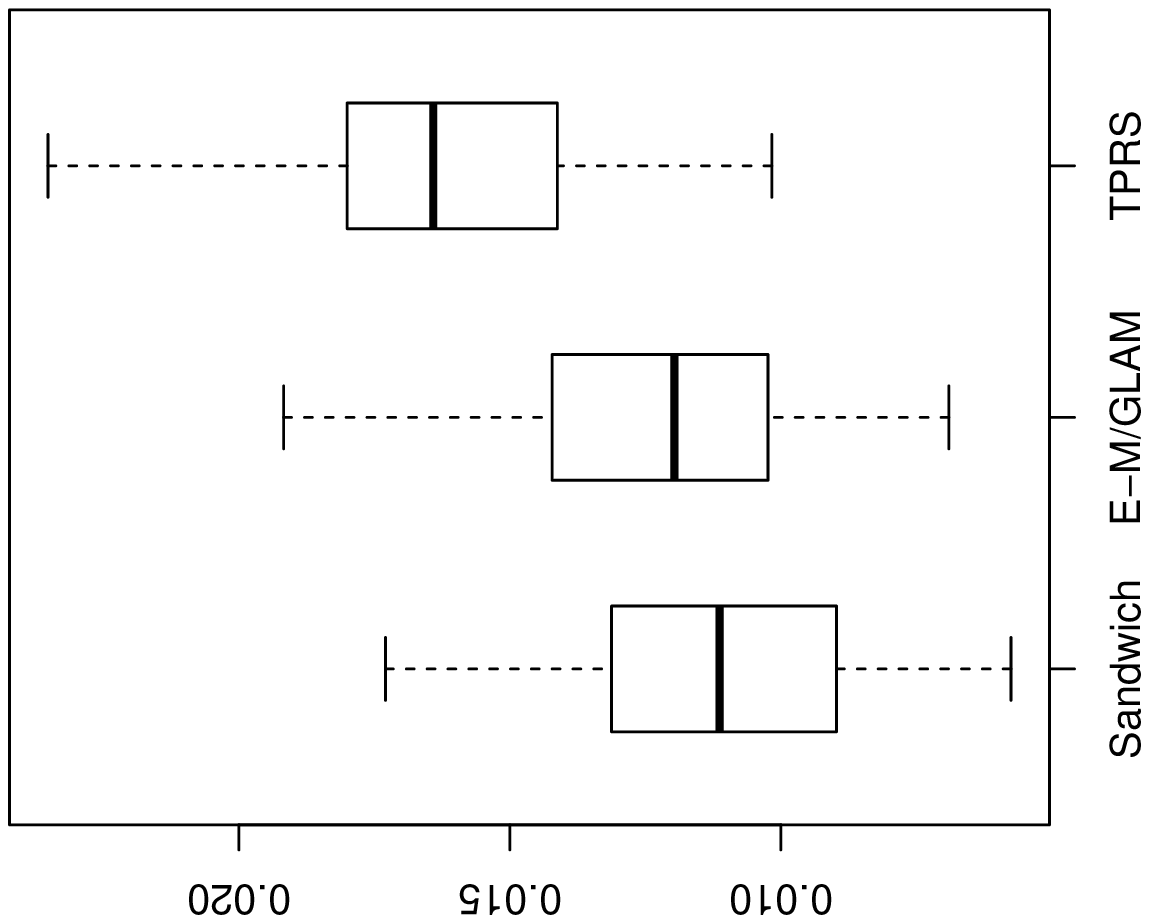}}\\
\subfigure[$f_2,\sigma=0.1$]{\includegraphics[ height=2in, angle=270]{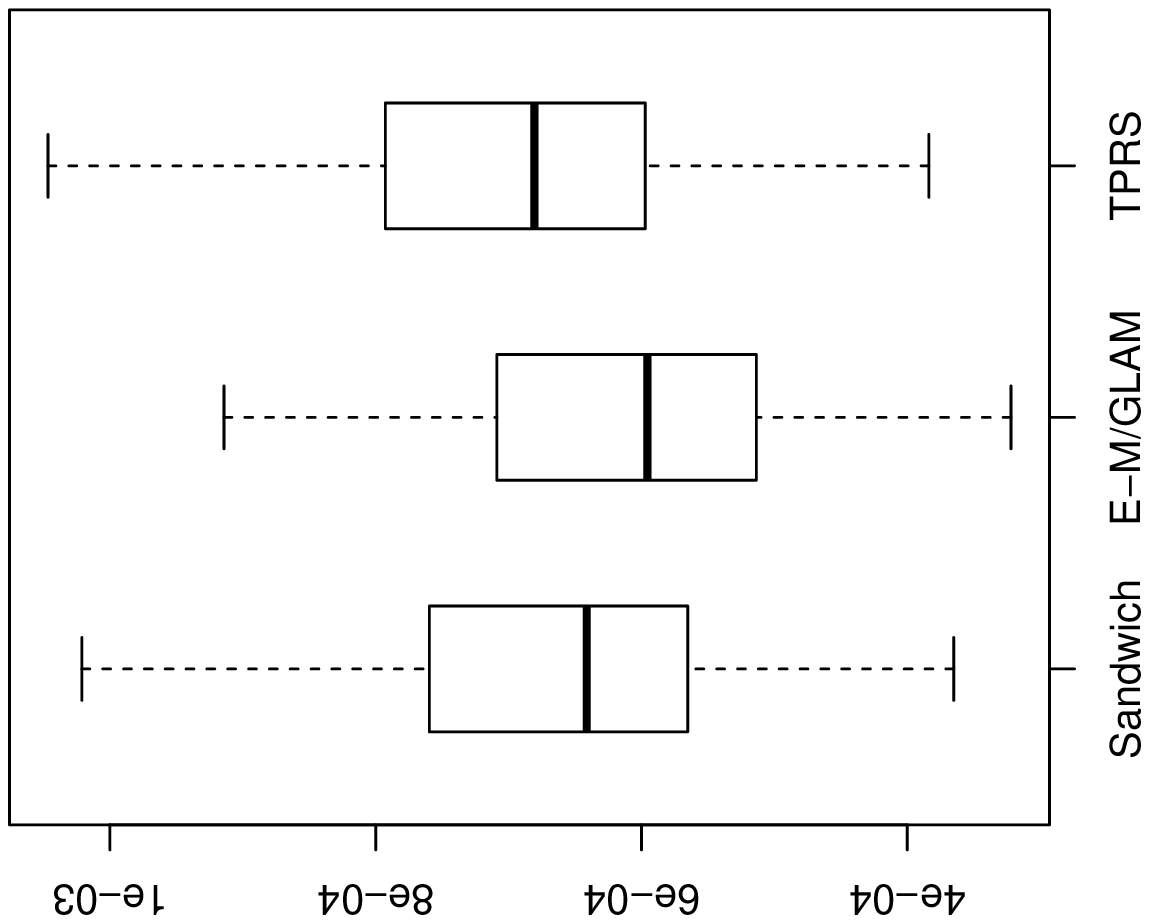}}
\subfigure[$f_2,\sigma=0.5$]{\includegraphics[height=2in, angle=270]{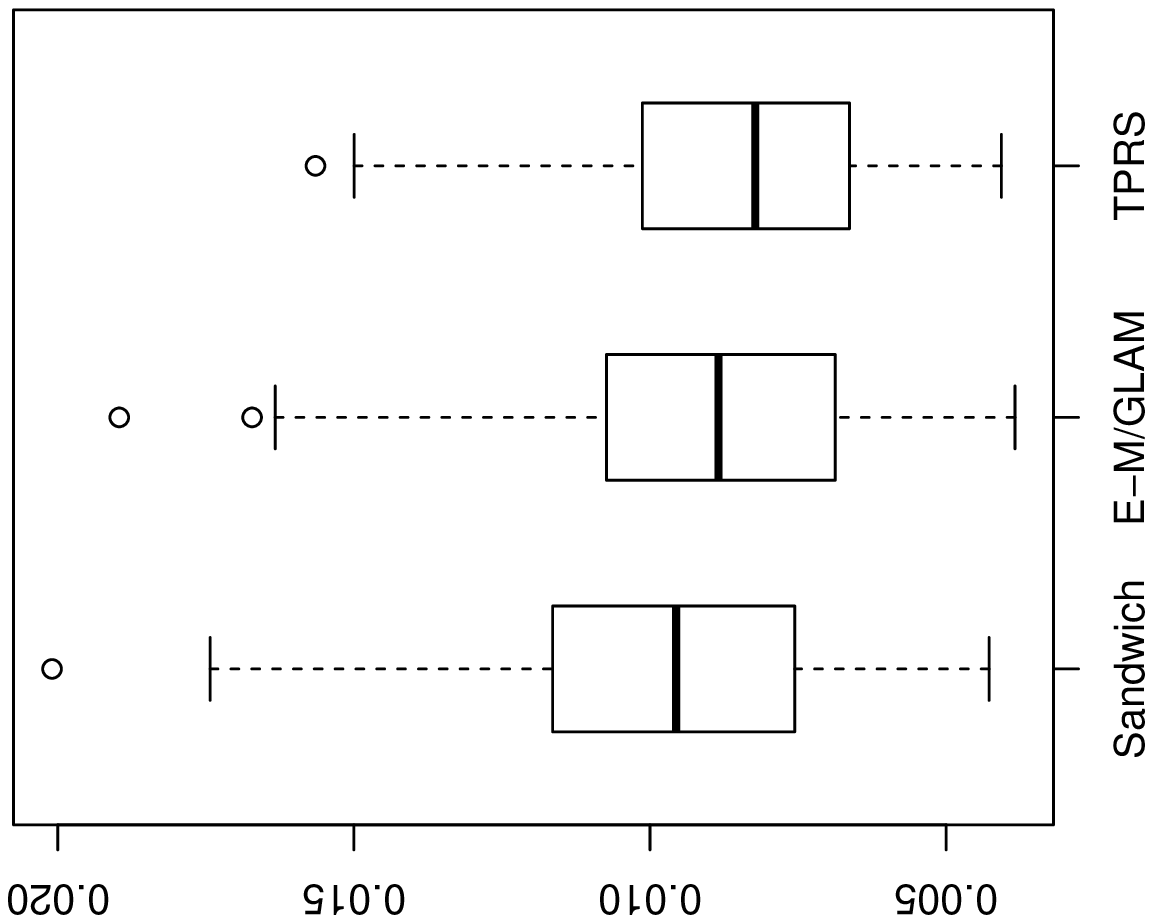}}\\
\end{center}
\caption{\label{fig1}Boxplots of the ISEs of three estimators for small samples}
\end{figure}

Performances of the three smoothers  were assessed at two sample sizes.  In the smaller sample study, each
test function was sampled on the $20\times 30$ regular grid on the unit square, and  random errors were iid $N(0,\sigma^2)$
with $\sigma$ equal to 0.1 and 0.5. In each case, $100$ replicate data sets were generated and, for each replicate data,  the test function
was fitted by the three estimators and the integrated squared error (ISE) was calculated. For the spline basis and knots settings, based on the recommendation in Remark~\ref{rem3},
$10$ and $15$  equidistant knots were used for the $x$- and $z$-axis for the two P-spline estimators. Thus, a total of 150 knots were used to construct the B-spline basis.
Cubic B-splines were used with a second order difference penalty. For the thin plate regression estimator (TPRS), we implemented the TPRS using the function ``bam" in a R package ``mgcv" developed by Simon Wood.  In this study, TPRS  was used with a rank of 150 (i.e., the basis dimension is 150). For all  three estimators, the smoothing parameters were chosen by \mbox{GCV}. The performances of the three estimators
were evaluated by the mean ISEs (MISEs; see Table~\ref{table1}) and also boxplots of the ISEs (see Figure~\ref{fig1}).
\begin{table}
\caption{MISEs of three estimators for a small sample size (data on a $20\times 30 $ grid).}
\centering
\fbox{%
\begin{tabular}{ccccc}
\\[1pt]
&$\sigma$ &  Sandwich smoother  & E-M/GLAM & TPRS\\\\[1pt] \hline
\\[1pt]
 \multirow{2}{*}{$f_1$}
&0.1&$8.13\times 10^{-4}$&$9.29\times 10^{-4}$&$1.46\times 10^{-3}$\\
&0.5&$1.08\times 10^{-2}$&$1.18\times 10^{-2}$&$1.56\times 10^{-2}$\\[1pt]\\
 \multirow{2}{*}{$f_2$}
&0.1&$6.45\times 10^{-4}$&$5.73\times 10^{-4}$&$6.68\times 10^{-4}$\\
&0.5&$9.25\times 10^{-3}$&$8.34\times 10^{-3}$&$8.06\times 10^{-3}$\\\\[1pt]
\end{tabular}}
\label{table1}
\end{table}

From Table \ref{table1} we can see that sandwich smoother did better than E-M/GLAM for estimating $f_1$ while E-M/GLAM was better for estimating $f_2$. The boxplots in Figure~\ref{fig1} show that the two P-spline methods are essentially comparable. Compared to the two P-spline methods, TPRS gave larger MISEs except for one case. One explanation for the relative inferior performance of TPRS for estimating $f_1$ is that TPRS is isotropic and has only a single smoothing parameter so that the same amount of smoothing is applied in both directions, which might be not appropriate for $f_1$ as $f_1$ is quite smooth in $x$ and varies rapidly in $z$ (see Figure~\ref{fig0}).

A larger sample simulation study with $n_1=60$ and $n_2=80$ was also done. For the two P-spline estimators, the numbers of knots were $K_1 =30$ and $  K_2 = 35$.  The rank of the TPRS was 1050, which was the total number of knots used in the two P-spline estimators.  All the other settings were the same as in the  smaller sample study. The resulting MISEs and boxplots gave the same conclusions as in the smaller sample study. To save space, we do not show the results here.

\subsection{Computation speed} \label{sec:speed}
The computation speed of the three spline smoothers for smoothing $f_2$  with varying numbers of data points was  assessed. For simplicity, we let $n_1=n_2$ and considered the case $\sigma=0.1$. We  selected the number of knots for the two P-spline smoothers  following the recommendation in Remark~\ref{rem3}.  We fixed the rank of TPRS to the total number of knots used in the P-spline smoothers. For the two P-spline smoothers, the computation times reported are for the case where the search for optimal smoothing parameters is over a $20\times 20$ log scale grid in $[-5,4]^2$.  A finer grid with $40^2$ grid points was also used. The computation was done on  2.83GHz computers running Windows  with 3GB of RAM. Table~\ref{time} summarizes the results and shows that the sandwich smoother is by far the fastest method. Note that the values in parenthesis are the computation time using the finer grid.

To further illustrate its computational capacity, the sandwich smoother was applied to large data with sizes of $300^2$ and $500^2$. For cubic B-splines coupled with second-order difference penalty, Theorem~\ref{thm1} suggested  choosing $K_1>n^{3/10}$ and $K_2>n^{3/10}$. So we let $K_1 = K_2$ with $K_1K_2$ close to $n^{3/5+0.1}$ in the simulations. We also evaluated the speed of E-M/GLAM. To save time, the E-M/GLAM was run for only $25$ pairs of smoothing parameters and the computation time  was multiplied by 16 (64) so as to be comparable to that of the sandwich smoother on the coarse (fine) grid.  The results in Table~\ref{time} show that the sandwich smoother could process large data quite fast on a personal computer while the E-M/GLAM is much slower. The TPRS was not applied to these large data as it would require more memory space than the computer could provide.

\begin{table}
\caption{\label{time}Computation time (in seconds) of three estimators averaged over 100 data sets on 2.83GHz computers running Windows with 3GB of RAM. The times for the sandwich smoother and E-M/GLAM are for a $20 \times 20$ grid of smoothing parameter values and (in parenthesis) for  a finer $40 \times 40$ grid.  For $n=20^2, 40^2$ and $80^2$, the number of knots for each axis is chosen by the recommendation in Remark~\ref{rem3}. For $n=300^2$ and $500^2$, the total number of knots for the sandwich smoother is approximately $n^{3/5 + 0.1}$ as suggested by Theorem~\ref{thm1}. }
\centering
\fbox{%
\begin{tabular}{ccclc}
\\[1pt]
$n$ & $K_1K_2$&  Sandwich smoother  & E-M/GLAM & TPRS\\\\[1pt]
 \hline
\\[1pt]
$20^2$&$10^2$&0.06(0.24)&4.09(19.74)&0.53\\
$40^2$&$20^2$&0.08(0.30)&94.76(344.13)&19.50\\
$80^2$&$35^2$&0.13(0.45)&1379.21(5487.33)&1032.07\\
$300^2$&$42^2$&0.18(0.58)&3798.23(15192.92)&--\\
$500^2$&$57^2$&0.32(0.89)&21023.44(84093.76)&--\\[1pt]
\end{tabular}}
\end{table}

To summarize, the simulation study  here and also the fast implementation in Section~\ref{sec:algorithm}  show the advantage of the sandwich smoother over the two other  estimators. So when computation time is of concern, the sandwich smoother might be preferred.

\section{Application: covariance function estimation}\label{sec:covariance}

As functional data analysis (FDA) has become a major research area, estimation of covariance functions has become an important application of bivariate smoothing.  Because functional data sets can be quite large, fast calculation of bivariate smooths is essential in FDA, especially when the bootstrap is used for inference. Local polynomial smoothing is a  popular method in estimating covariance
functions (see e.g., Yao {\it et al.}\ (2005) or Yao and Lee (2006)) while other smoothing methods such as kernel (Staniswalis and Lee, 1998) and  penalized splines (Di {\it et al.}, 2009) have  also been used.  In this section, through a simulation study we compare the performance of the sandwich smoother and local polynomials for estimating a covariance function when the data
are observed or measured at a fixed grid.

Let $\{X(t): t\in [0,1]\}$ be a stochastic process with a continuous covariance function $K(s,t) = \textrm{cov}\{X(s), X(t)\}$.
For simplicity, we assume $\textrm{E}X(t) = 0, t\in [0,1]$. Suppose $\{X_i(t), i = 1,\dots, n\}$ is a collection of independent realizations of the above stochastic process and we observe the
random functions $X_i$ at discrete design points with measurement errors,
$$
Y_{ij} = X_i(t_j) +\epsilon_{ij}, 1\leq j\leq J, 1\leq i\leq n,
$$
where $J$ is the number of measurements per curve,  $n$ is the total number of curves, and the $\epsilon_{ij}$ are i.i.d.\ measurement errors with mean zero and finite variance and they are independent of the random functions $X_i$. Let $\mathbf{Y}_i = (Y_{i1},\dots, Y_{iJ})^T$.  An estimate of the covariance function can be obtained through smoothing the sample covariance matrix $ n^{-1}\sum_{i=1}^n \mathbf{Y}_i\mathbf{Y}_i^T$ by a bivariate smoother. Because we are smoothing a symmetric matrix, for the sandwich smoother we use two identical univariate smoother matrices so there is only one smoothing parameter to select. We use the commonly used local linear smoother (Yao {\it et al.}, 2005, Hall {\it et al.}, 2006) for comparison and the bandwidth is selected by the leave-one-curve-out cross validation. We wrote our own {\tt R} implementation of the estimator used by Yao {\it et al.}\ (2005), since their code is in Matlab.

\begin{figure}[htp]
\centering
\includegraphics[width=5in, angle=270]{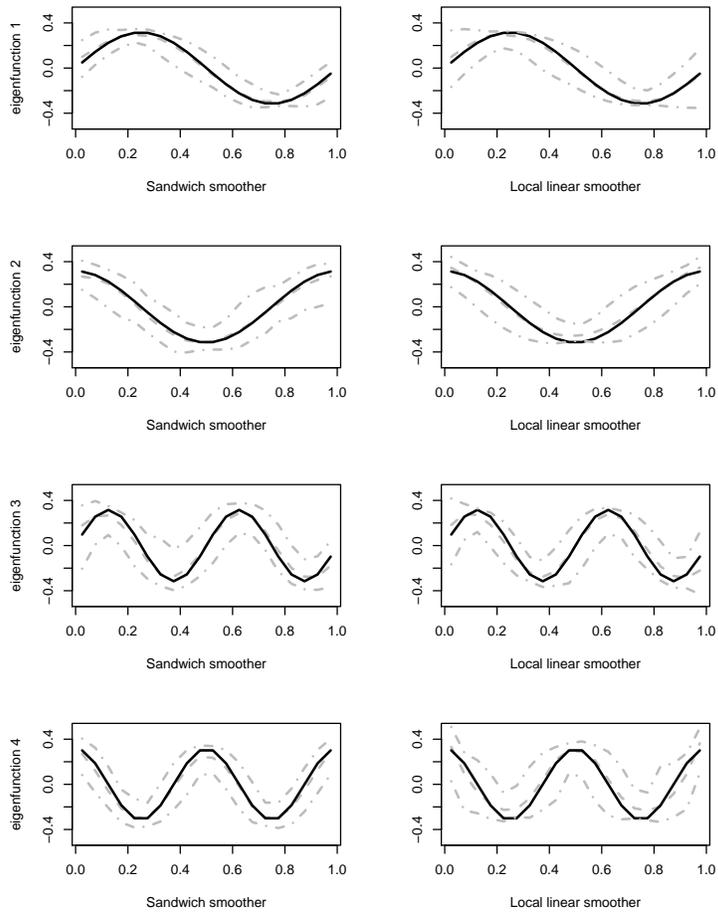}
\caption{\label{fig3}True and estimated eigenfunctions  replicated 100 times with $(n,J) = (25,20)$ for case 1. The variance of noises is $0.25$. Each box shows the true eigenfunction (solid black lines), the pointwise median estimated eigenfunction (dashed gray lines),  the 5th and 95th pointwise percentile curves (dot-dashed gray  lines). The left column is for the sandwich smoother and the right one is for local linear smoother.}
\end{figure}

We let $K(s,t) = \sum_{k=1}^4 \lambda_k \psi_k(s)\psi_k(t)$ where the eigenvalues $\lambda_k = 0.5^{k-1}, k = 1, 2, 3, 4$,  and $\{\psi_1,\dots, \psi_4\}$ are the eigenfunctions   from either of the following
\begin{eqnarray*}
\text{Case 1:}& \left\{\sqrt{2}\sin(2\pi t),\sqrt{2}\cos(2\pi t),\sqrt{2}\sin(4\pi t),\sqrt{2}\cos(4\pi t)\right\},\\
\text{Case 2:}& \left\{1, \sqrt{3}(2t-1),\sqrt{5}(6t^2-6t+1),\sqrt{7}(20t^3-30t^2+12t-1)\right\}.
\end{eqnarray*}
The above two sets of eigenfunctions were used in Di {\it et al.}\ (2009), Greven {\it et al.}\ (2010),  and Zipunnikov {\it et al.}\ (2011). We let $\sigma = 0.5$. We simulate 100 datasets and evaluate the two bivariate smoothers in terms of mean ISEs (MISEs). The results are given in Table~\ref{table2}.  From Table~\ref{table2}, for case 1 with $(n,J) = (25,20)$ the local linear smoother  is slightly better with smaller mean and standard deviation of ISE's and for other cases the two smoothers give close results. The estimated  eigenfunctions by the two smoothers for case 1 with $(n,J) = (25,20)$ are shown  in Figure~\ref{fig3}. The figure shows that both smoothers  estimate the eigenfunctions well. We found similar results for $(n,J) = (100,40)$ (results not shown).

\begin{table}
\caption{\label{table2} MISEs of the sandwich smoother and the local linear  smoother for estimating a covariance function. The number in parenthesis is the standard deviation of ISE's.}
\centering
\fbox{%
\begin{tabular}{cccc}
\\[1pt]
$(n,J)$ &  Case & Sandwich smoother & Local linear smoother\\[1pt] \hline
\\[1pt]
 \multirow{2}{*}{$(25, 20)$}
&$1$&$.053(.035)$&$.050(.026)$\\
&$2$&$.199(.139)$&$.204(.144)$\\[1pt]\\
 \multirow{2}{*}{$(100, 40)$}
&$1$&$.014(.008)$&$.013(.008)$\\
&$2$&$.050(.034)$&$.050(.036)$\\\\[1pt]
\end{tabular}}
\end{table}

We also compared the computation time of the two smoothers using case 1 for various values of $J$. For the sandwich smoother, we searched over twenty smoothing parameters. For the local linear smoother, we fixed the bandwidth. Note that selecting the bandwidth by the leave-one-curve-out cross validation means the computation time of the local linear smoother will be multiplied by the number of bandwidths and also the number of curves.   Table~\ref{table3} shows that the sandwich smoother is  much faster to compute than the local linear smoother for covariance function estimation even when the bandwidth for the latter is fixed.

\begin{table}
\caption{\label{table3}Computation time (in seconds) for smoothing an $J \times J$ covariance matrix using the sandwich smoother and the local linear smoother.  With one exception, the computation times are averaged over 100 data sets on 2.83GHz computers running Windows with 3GB of RAM. The number of curves  is fixed at 100. The bandwidth for the local linear smoother is fixed in the computations. The exception is that the computation time for the local linear smoother when $J=320$ is averaged over 10 datasets only. }
\centering
\fbox{%
\begin{tabular}{ccc}
\\[1pt]
$J$ &  Sandwich smoother  & Local linear smoother \\\\[1pt]
 \hline
\\[1pt]
$40$&0.02&2.98\\
$80$&0.03&50.04\\
$160$&0.05&961.42\\
$320$&0.16&13854.40\\[1pt]
\end{tabular}}
\end{table}

To summarize, the simulation study suggests that  for covariance function estimation when functional data are measured at a fixed grid, the sandwich smoother is comparable to the local linear smoother in terms of MISEs. The sandwich smoother is considerably faster to compute than the local linear smoother.

\section{Multivariate P-splines}\label{sec:multi}
We extend the sandwich smoother to array data of dimensions greater than two. Suppose we have a  nonparametric regression model with $d\geq 3$ covariates
\[
y_{i_1,\dots, i_d} = \mu(x_{i_1},\dots, x_{i_d}) +\epsilon_{i_1,\dots, i_d},\quad 1\leq i_k\leq n_k, 1\leq k\leq d,
\]
so the data are collected on a $d$-dimensional grid. For simplicity, assume the covariates are in $[0,1]^d$. As in the bivariate case, we  model the $d$-variate function $\mu(x_1,\dots, x_d)$ by tensor-product B-splines of $d$ variables
$
\sum_{\kappa_1,\kappa_2, \dots, \kappa_d} \theta_{\kappa_1,\kappa_2,\dots,\kappa_d}B_{\kappa_1}^1(x_1)B_{\kappa_2}^1(x_2)\cdots B_{\kappa_d}^d(x_d),
$
where $B_{\kappa_1}^1, B_{\kappa_2}^2, \dots, B_{\kappa_d}^d$ are the B-spline basis functions.   We smooth along all  covariates simultaneously so that the fitted values and the data satisfy
\begin{equation}
\label{multi-haty}
\hat{\mathbf{y}} = \left(\mathbf{S}_d\otimes \mathbf{S}_{d-1}\otimes\dots \otimes \mathbf{S}_1\right)\mathbf{y},
\end{equation}
where $\mathbf{S}_i$ is the smoother matrix for the $i$th covariate using  P-splines as in~(\ref{S}), $\mathbf{y}$ is the data vector organized first by $x_1$, then by $x_2$,  and so on, and $\hat{\mathbf{y}}$ is organized the same way as $\mathbf{y}$.
Similar to equation~(\ref{eqnobj}), the estimate of coefficients $\hat{\boldsymbol{\theta}}$ satisfies
\[
\left(\mathbf{\Lambda}_d\otimes\mathbf{\Lambda}_{d-1}\otimes\cdots\otimes \mathbf{\Lambda}_1\right)\hat{\boldsymbol{\theta}}=  (\mathbf{B}_d\otimes\mathbf{B}_{d-1}\otimes\cdots\otimes \mathbf{B}_1)^T\mathbf{y},
\]
and the penalized estimate is
\[
\hat{\mu}(x_1,x_2,\dots, x_d)=
\sum_{\kappa_1,\kappa_2, \dots, \kappa_d}\hat{\theta}_{\kappa_1,\kappa_2,\dots,\kappa_d}B_{\kappa_1}^1(x_1)B_{\kappa_2}^1(x_2)\cdots B_{\kappa_d}^d(x_d).
\]
\subsection{Implementation of the multivariate P-splines} \label{sec:multi:algorithm}

 Two computational issues occur for smoothing data on a multi-dimensional grid. The first issue is that unless the sizes of $\mathbf{S}_i$'s are all small, the storage and computation of $\mathbf{S}_d\otimes \mathbf{S}_{d-1}\otimes\dots \otimes \mathbf{S}_1$ will be challenging. The second issue is selection of smoothing parameters.  Because of the large number of smoothing parameters involved, finding the smoothing parameters that minimize some model selection criteria such as GCV can be difficult.

The generalized linear array model by Currie {\it et al.}\,(2006) provided an elegant solution to the first issue by making use of the array structures of the model matrix as well as the data. The smoother matrix $\mathbf{S}_d\otimes \mathbf{S}_{d-1}\otimes\dots \otimes \mathbf{S}_1$ in multivariate smoothing has a tensor product structure, hence $\hat{\mathbf{y}}$ in~(\ref{multi-haty}) can be computed efficiently by a sequence of nested operations on $\mathbf{y}$ by  the \text{GLAM} algorithm. For instance,  consider $d=3$. Then $\hat{\mathbf{y}}$ can be computed efficiently
with one line of R code:
\begin{verbatim}
# The function "RH" is the rotated H-transform of an array by a matrix
# see Currie et al. (2006)
yhat = as.vector(RH(S3,RH(S2,RH(S1,Y))))
\end{verbatim}

\noindent We wrote an {\tt R} version of the {\tt RH} function.

The second issue can be easily handled for the multivariate fast P-splines. Because of the tensor product structure of the smoother matrix, the fast implementation in Section~\ref{sec:algorithm} can be  generalized for the multivariate case.   As an illustration, we show how to compute the trace of the smoother matrix.  We first compute the singular value decompositions for all $\mathbf{S}_i$ so that~(\ref{trace2}) holds for all $i=1,\dots, d$, then we compute the trace of the smoother matrix  by
\[
\tr\left(\mathbf{S}_d\otimes \mathbf{S}_{d-1}\otimes\dots \otimes \mathbf{S}_1\right) = \prod_{i=1}^d \tr(\mathbf{S}_i)
\]
using the identity in~(\ref{trace}) repeatedly. Note that $\tr(\mathbf{S}_i)$ has a similar expression as in~(\ref{trace2}) for all $i$.

The sandwich smoother does not have a GLM weight matrix and when it is used for bivariate smoothing, there is no need for rotation of arrays, so we do not consider the bivariate sandwich smoother to be a GLAM algorithm. However, our implementation for the bivariate sandwich smoother makes use of tensor product structures to simplify calculations similar to what the GLAM does.

\subsection{An example}
Smoothing simulated image data of size $128\times 128\times 24$ with a $20^3$ grid of smoothing parameters, the sandwich smoother takes about 20 seconds on a 2.4GHz computer running Mac software with 4GB of RAM. We have not found the computation time of other smoothers, but we can give a crude lower bound.  We see in Table \ref{time} that E-M/GLAM takes about 1400 seconds (over 20 minutes) on a $80^2$ two-dimensional grid where the smoothing parameters are searched over  a $20\times 20$ grid. Searching over a $20\times 20\times 20$ grid to select the smoothing parameters, the number of times of  GCV computation is now 20 times more. Moreover, for each GCV computation, E-M/GLAM will need much more time for smoothing data of size  $128\times 128\times 24$ which is much larger. Therefore, the  E-M/GLAM estimator's computation time for smoothing a $128\times 128\times 24$ will be many hours for an algorithm that does not compute GCV as efficiently as the sandwich smoother does.

\section*{Acknowledgement}

This research was partially supported by National Science Foundation grant DMS-0805975 and National Institutes of Health grant R01-NS060910. Luo Xiao's research was partly supported by the National Center for Research Resources Grant UL1-RR024996.
We thank Professor Iain Currie for his helpful discussion on the GLAM algorithm. We  thank the two referees and the associate editor for their most helpful comments and suggestions which greatly improve this paper. We are grateful to one referee who suggested the name ``sandwich smoother".

\appendix

\section{Appendix: Derivation of equation~(\ref{gcv0}) } \label{sec:derivation}
First we have
  \begin{equation*}
  \|\hat{\mathbf{Y}}-\mathbf{Y}\|_F^2 = (\hat{\mathbf{y}} -\mathbf{y})^T(\hat{\mathbf{y}} -\mathbf{y})= \hat{\mathbf{y}}^T\hat{\mathbf{y}} - 2 \hat{\mathbf{y}}^T\mathbf{y} + \mathbf{y}^T\mathbf{y}.
\end{equation*}
It can be shown by~(\ref{gcv01}) that
\begin{equation*}
\begin{split}
\hat{\mathbf{y}}^T\hat{\mathbf{y}} &=\tilde{\mathbf{y}}^T(\boldsymbol{\Sigma}_2\otimes\boldsymbol{\Sigma}_1) (\mathbf{A}_2\otimes\mathbf{A}_1)^T(\mathbf{A}_2\otimes\mathbf{A}_1)(\boldsymbol{\Sigma}_2\otimes\boldsymbol{\Sigma}_1)\tilde{\mathbf{y}}\\
& =\tilde{\mathbf{y}}^T(\boldsymbol{\Sigma}_2\otimes\boldsymbol{\Sigma}_1)(\boldsymbol{\Sigma}_2\otimes\boldsymbol{\Sigma}_1)\tilde{\mathbf{y}}\\
& = | \tilde{\mathbf{y}}^T(\boldsymbol{\Sigma}_2\otimes\boldsymbol{\Sigma}_1)|^2\\
&= \left\{\tilde{\mathbf{y}}^T\left(\tilde{\mathbf{s}}_2\otimes\tilde{\mathbf{s}}_1\right)\right\}^2.
\end{split}
\end{equation*}
In the above derivation, $|\cdot|$ denotes the Euclidean norm in the second to last equality; we used the facts that  $\mathbf{A}_i^T\mathbf{A}_i =\mathbf{I}_{c_i}$ and that both $\boldsymbol{\Sigma}_2$ and $\boldsymbol{\Sigma}_1$ are diagonal matrices. Similarly we obtain
$$\hat{\mathbf{y}}^T\mathbf{y} = \left\{ \tilde{\mathbf{y}}^T \left(\tilde{\mathbf{s}}_2^{1/2}\otimes\tilde{\mathbf{s}}_1^{1/2}\right)\right\}^2
 $$
and hence establishes~(\ref{gcv0}).

\section{Appendix: Proof of theorems} \label{sec:proof}

\begin{lem}\label{H_m} 
The univariate kernel function $H_m(x)$ defined in~(\ref{kernel}) satisfies the following:
\begin{equation*}
\int_{-\infty}^{\infty} x^{l}H_m(x)\,\mathrm{d}x =
\begin{cases}
1&:\quad l=0\\
0&: \quad l\quad\mbox{is odd}\\
 0 &:\quad l \quad\mbox{is even and }\quad 2\leq l\leq 2m-2\\
(-1)^{m+1} (2m)! &:\quad l=2m
\end{cases}.
\end{equation*}
Hence $H_m(x)$ is of order $2m$.
\end{lem}
{\it Proof of Lemma \ref{H_m}:} We need to calculate two types of integrals  $\int x^{l} \exp(ax) \cos(bx)\,\mathrm{d}x$
and $\int x^{l} \exp(ax) \sin(bx)\,\mathrm{d}x$. Those indefinite integrals are given by results 3 and 4  in Gradshteyn and Ryzhik (2007, pp.\ 230).
Then a routine calculation gives the desired result. Part of the lemma  is derived in Wang {\it et al.}\ (2011). Details of derivation can be found in
 Xiao {\it et al.}\ (2011).

Before proving Proposition~\ref{prop1}, we need the following lemma:
\begin{lem}\label{lem1}
Use the same notation in Proposition~\ref{prop1} and assume all conditions and assumptions in Proposition~\ref{prop1} are satisfied.  For $(x,z)\in
(0,1)\times (0,1)$, there exists a constant $C>0$   such that
$$
\hat{\mu}(x,z)=\sum_{i,j} y_{i,j}\left[\left\{\sum_{\kappa,r}B_{\kappa}^1(x)
B_r^1(x_i)S_{\kappa,r,x}\right\}\left\{\sum_{\ell,s}
B_{\ell}^2(z)B_s^2(z_j)S_{\ell,s,z}\right\} + \tilde{b}_{i,j}(x,z) \right],
$$
where $\tilde{b}_{i,j}(x,z) = O\left[\exp\left\{-C\min(h_{n,1}^{-1}, h_{n,2}^{-1})\right\}\right]$.
\end{lem}
{\it Proof of Lemma~\ref{lem1}}: By~(\ref{est}), $\hat{\mu}(x,z) = \sum \hat{\theta}_{\kappa,\ell} B_{\kappa}^1(x) B_{\ell}^2(z)$.  We only need to consider $\hat{\theta}_{\kappa,\ell}$ for which $B_{\kappa}^1(x)$ and $B_{\ell}^2(z)$ are both non-zero. Hence assume $\kappa$ and $\ell$ satisfy $\kappa\in (K_1x-p_1-1, K_1 x+p_1+1)$, $\ell\in (K_2 z-p_2-1,K_2z+p_2+1)$. Let $q_1 = \max(p_1,m_1)$ and $q_2 = \max(p_2,m_2)$.
Denote by $\mathbf{\Lambda}_{1,j}$  the $j$th column of $\mathbf{\Lambda}_1$ and $\mathbf{\Lambda}_{2,j}$ the $j$th column of $\mathbf{\Lambda}_2$.
As shown in Xiao {\it et al.}\ (2011) and  Li and Ruppert (2008), there exist vectors $\mathbf{S}_{\kappa,x}$ and a constant $C_3>0$ so that for $q_1<j<c_1-q_1$, $\mathbf{S}_{\kappa,x}^T\mathbf{\Lambda}_{1,j}  = \delta_{\kappa,j}$, and for $1\leq j\leq q_1$ or  $c_1-q_1\leq j\leq c_1$,
$\mathbf{S}_{\kappa,x}^T\mathbf{\Lambda}_{1,j}  = O\left[\exp\left\{-C_3 h_{n,1}^{-1}\min(x,1-x)\right\}\right]$. Here $\delta_{\kappa,j} = 1$ if $j = \kappa$ and $0$ otherwise. Similarly, there exist vectors $\mathbf{S}_{\ell,z}$ and a constant $C_4>0$ such that for $q_2<j<c_2-q_2$, $\mathbf{S}_{\ell,z}^T\mathbf{\Lambda}_{2,j}  = \delta_{\ell,j}$, and for $1\leq j\leq q_2$ or  $c_2-q_2\leq j\leq c_2$, $\mathbf{S}_{\ell,z}^T\mathbf{\Lambda}_{2,j}  = O\left[\exp\left\{-C_4h_{n,2}^{-1}\min(z,1-z)\right\}\right]$.  Let $\tilde{\theta}_{\kappa,\ell} = \left(\mathbf{S}_{\ell,z}\otimes \mathbf{S}_{\kappa,x}\right)^T\left(\mathbf{\Lambda}_2 \otimes \mathbf{\Lambda}_1\right)\hat{\boldsymbol{\theta}}$ and $C = \min\left\{C_3\min(x,1-x), C_4\min(z,1-z) \right\}$, then
\begin{equation}
\label{tilde_theta}
\tilde{\theta}_{\kappa,\ell} - \hat{\theta}_{\kappa,\ell} = \sum_{i,j} \tilde{b}_{i,j,\kappa,\ell} y_{i,j},
\end{equation}
where $\tilde{b}_{i,j,\kappa,\ell} = O\left[\exp\left\{-C\min(h_{n,1}^{-1}, h_{n,2}^{-1} )\right\}\right]$. By equation~(\ref{eqnobj}),
\[
\tilde{\theta}_{\kappa,\ell} =  \left(\mathbf{S}_{\ell,z}\otimes \mathbf{S}_{\kappa,x}\right)^T\left(\mathbf{B}_2^T\otimes \mathbf{B}_1^T\right) \mathbf{y} =\left(\mathbf{S}_{\ell,z}^T\mathbf{B}_2^T\otimes \mathbf{S}_{\kappa,x}^T\mathbf{B}_1^T\right)\mathbf{y}
 = \mathbf{S}_{\kappa,x}^T\left(\mathbf{B}_1^T\mathbf{Y}\mathbf{B}_2\right)\mathbf{S}_{\ell,z}.
\]
Letting $S_{\kappa,r,x}$ be the $r$th element of $\mathbf{S}_{\kappa,x}$ and similarly  $S_{\ell,s, z}$  the $s$th element of $\mathbf{S}_{\ell,z}$,  we express $\tilde\theta_{\kappa,\ell}$ as
a double sum
\begin{equation}
\label{tilde_theta2}
\tilde \theta_{\kappa,\ell} =  \sum_{r,s}S_{\kappa,r,x}\left\{\sum_{i,j}B_r^1(x_i)y_{i,j}B_s^2(z_j)\right\}S_{\ell,s,z}
=  \sum_{i,j}y_{i,j}\left\{\sum_r
B_r^1(x_i)S_{\kappa,r,x}\right\}\left\{\sum_s
B_s^2(z_j)S_{\ell,s,z}\right\}.
\end{equation}
With equations~(\ref{est}),~(\ref{tilde_theta}) and~(\ref{tilde_theta2}), we have
\begin{equation*}
\begin{split}
\hat{\mu}(x,z) &= \sum_{\kappa,\ell} \tilde \theta_{\kappa,\ell} B_{\kappa}^1(x)B_{\ell}^2(z) + \sum_{\kappa,\ell} (\hat\theta_{\kappa,\ell}-\tilde\theta_{\kappa,\ell}) B_{\kappa}^1(x)B_{\ell}^2(z)\\
&=\sum_{i,j} y_{i,j}\left[\left\{\sum_{\kappa,r}B_{\kappa}^1(x)
B_r^1(x_i)S_{\kappa,r,x}\right\}\left\{\sum_{\ell,s}
B_{\ell}^2(z)B_s^2(z_j)S_{\ell,s,z}\right\} + \tilde{b}_{i,j}(x,z) \right],
\end{split}
\end{equation*}
where $\tilde{b}_{i,j}(x,z) = O\left[\exp\left\{-C\min(h_{n,1}^{-1}, h_{n,2}^{-1})\right\}\right]$.

{\it Proof of Proposition~\ref{prop1}}:  Let $\tilde{\lambda}_1=\lambda_1 K_1 n_1^{-1}=(K_1h_{n,1})^{2m_1}$ and $\tilde{\lambda}_2=\lambda_2 K_2n_2^{-1}=(K_2h_{n,2})^{2m_2}$. By Proposition 5.1 in Xiao {\it et al.}\ (2011), there exists some constants $0<\phi_1, \phi_2<\infty$ such that
\begin{equation}
\label{kernel_H1}
\begin{split}
&n_1h_{n,1}\sum_{k,r}B_k^1(x)B_r^1(x_i)S_{k,r,x}\\
=&H_{m_1}\left(\frac{|x-x_i|}{h_{n,1}}\right)+
\delta_{\{p_1>m_1\}}\left[O\left(\tilde \lambda_1^{-2+\frac{1}{2m_1}}\right)+ \delta_{\{|x-x_i|< \phi_1/K_1\}}O\left(\tilde \lambda_1^{-\frac{p_1}{p_1-m_1}+\frac{1}{2m_1}}\right)\right]\\
&+\exp\left(-\phi_2 \frac{|x-x_i|}{h_{n,1}}\right)\left[O\left(\tilde{\lambda}_1^{-\frac{1}{m_1}}\right)+\delta_{\{m_1=1\}}\delta_{\left\{|x-x_i|\leq (p_1+1)\tilde \lambda_1^{-1/(2m_1)}\right\}}O\left(\tilde{\lambda}_1^{-\frac{1}{2m_1}}\right)\right].
\end{split}
\end{equation}
Here $\delta_{\{p_1>m_1\}}=1$ if $p_1>m_1$ and 0 otherwise; the other $\delta$ terms are similarly defined.
Similarly, there exist some constants $0<\phi_3,\phi_4<\infty$ such that
\begin{equation}
\label{kernel_H2}
\begin{split}
&n_2h_{n,2}\sum_{\ell,s}B_{\ell}^2(z)B_s^2(z_j)S_{\ell,s,z}\\
=&H_{m_2}\left(\frac{|z-z_j|}{h_{n,2}}\right)+
\delta_{\{p_2>m_2\}}\left[O\left(\tilde \lambda_2^{-2+\frac{1}{2m_2}}\right)+ \delta_{\{|z-z_j|< \phi_3/K_2\}}O\left(\tilde \lambda_2^{-\frac{p_2}{p_2-m_2}+\frac{1}{2m_2}}\right)\right]\\
&+ \exp\left(-\phi_4\frac{|z-z_j|}{h_{n,2}}\right)\left[O\left(\tilde{\lambda}_2^{-\frac{1}{m_2}}\right)+\delta_{\{m_2=1\}}\delta_{\left\{|z-z_j|\leq (p_2+1)\tilde \lambda_2^{-1/(2m_2)}\right\}}O\left(\tilde{\lambda}_2^{-\frac{1}{2m_2}}\right)\right].
 \end{split}
\end{equation}
Let
\begin{align*}
d_{i,1} &= \sum_{k,r}B_k^1(x)B_r^1(x_i)S_{k,r,x}- (n_1h_{n,1})^{-1} H_{m_1}\left\{h_{n,1}^{-1}(x-x_i)\right\},\\
d_{i,2} &= \sum_{\ell,s}B_{\ell}^2(z)B_s^2(z_j)S_{\ell,s,z}- (n_2h_{n,2})^{-1}H_{m_2}\left\{h_{n,2}^{-1}(z-z_j)\right\},\\
b_{i,j}(x,z) &=\frac{1}{n_1h_{n,1}}H_{m_1}\left(\frac{|x-x_i|}{h_{n,1}}\right)d_{i,2} + \frac{1}{n_2h_{n,2}}H_{m_2}\left(\frac{|z-z_j|}{h_{n,2}}\right) d_{i,2} + d_{i,1}d_{i,2} +\tilde{b}_{i,j}(x,z).
\end{align*}
It follows from Lemma~\ref{lem1} that $\hat\mu(x,z)-\mu^{\ast}(x,z) = \sum_{i,j} b_{i,j}(x, z)y_{i,j}$. Hence $\textrm{E}\{\hat\mu(x,z)-\mu^{\ast}(x,z)\}=\sum_{i,j} b_{i,j}(x,z) \mu(x_i,z_j)$ and $\textrm{var}\{\hat\mu(x,z)-\mu^{\ast}(x,z)\}=\sum_{i,j} b_{i,j}^2(x,z) \sigma^2(x_i,z_j)$.

To simplify notation, denote $\max\{(K_1h_{n,1})^{-2}, (K_2h_{n,2})^{-2}\}$ by $\xi$.  We prove $\textrm{E}\{\hat\mu(x,z)-\mu^{\ast}(x,z)\} =O(\xi)$  by showing that $\sum_{i,j} |b_{i,j}(x,z)\mu(x_i,z_j)|$ is  $O(\xi)$. By Lemma~\ref{lem1}, $\tilde{b}_{i,j}(x,z) = O\left[\exp\left\{-C\min(h_{n,1}^{-1}, h_{n,2}^{-1})\right\}\right]$. Since $h_{n,1} = O(n^{-\nu_1})$ and $h_{n,2} = O(n^{-\nu_2})$, $\tilde b_{i,j}(x,z) = n^{-1}o(\xi) $ and hence  $\sum_{i,j} |\tilde{b}_{i,j}(x,z)\mu(x_i,z_j)| = o(\xi).$ For simplicity, we shall only show that
\begin{equation}
\label{proof_eqn_1}
\sum_{i,j}\left | \frac{1}{n_1h_{n,1}}H_{m_1}\left(\frac{|x-x_i|}{h_{n,1}}\right)d_{i,2} \mu(x_i,z_j)\right| = O(\xi),
\end{equation}
and we use the case when $p_2\leq m_2$ as an example. Because
\begin{align*}
&\frac{1}{nh_n}\sum_{i,j}\left |H_{m_1}\left(\frac{|x-x_i|}{h_{n,1}}\right)\exp\left(-\phi_4\frac{|z-z_j|}{h_{n,2}}\right) \mu(x_i,z_j)\right| = O(1),\\
&\frac{1}{nh_n}\sum_{i,j}\left |H_{m_1}\left(\frac{|x-x_i|}{h_{n,1}}\right)\exp\left(-\phi_4\frac{|z-z_j|}{h_{n,2}}\right) \delta_{\left\{|z-z_j|\leq (p_2+1) \tilde{\lambda}_2^{-1/(2m_2)}\right\}}\mu(x_i,z_j)\right| =O\left\{\tilde \lambda_2^{-\frac{1}{2m_2}}\right\},
\end{align*}
and $\tilde\lambda_2^{-1/m_2} = (K_2h_{n,2})^{-2}$, equality~(\ref{proof_eqn_1}) is proved. The case when $p_2>m_2$ and the desired results involving $d_{i,1}$ can be similarly proved.

Next we show that $\textrm{var}\{\hat\mu(x,z)-\mu^{\ast}(x,z)\} = o\{(nh_n)^{-1}\}$, i.e., $\sum_{i,j} b_{i,j}^2(x,z) \sigma^2(x_i,z_j) = o\{(nh_n)^{-1}\}$. Note that $b_{i,j}^2(x,z)\sigma^2(x_i,z_j)$ can be expanded into a  sum of individual terms. With similar analysis as before, for each individual term in $b_{i,j}^2(x,z)\sigma^2(x_i,z_j)$, the double sum over $i,j$ is either $O\{(nh_n)^{-1} \tilde\lambda_1^{-2/m_1}\}$, $O\{(nh_n)^{-1} \tilde \lambda_2^{-2/m_2}\}$, or is of smaller order.


\textit{Proof of Theorem~\ref{thm1}}:
Proposition~\ref{prop1} states that the sandwich smoother is asymptotically equivalent to a kernel regression estimator with a product kernel  $H_{m_1}(x) H_{m_2}(z)$. To determine the asymptotic bias and variance of the  kernel estimator,
we conduct a similar analysis of multivariate kernel density estimator as in Wand and Jones (1995).  By Proposition~\ref{prop1},
\begin{equation}
\label{kernelest2}
\textrm{E}\{\hat{\mu}(x,z)\}= \frac{1}{nh_{n,1}h_{n,2}}\sum_{i,j}\mu(x_i,z_j)H_{m_1}\left(\frac{x-x_i}{h_{n,1}}\right)H_{m_2}\left(\frac{z-z_j}{h_{n,2}}\right)+O(\xi),
\end{equation}
where we continue using the notation $\xi = \max\{(K_1h_{n,1})^{-2}, (K_2h_{n,2})^{-2}\}$. Let
\begin{equation}
\begin{split}
\mu_0(x,z)&=\frac{1}{nh_{n,1}h_{n,2}}\sum_{i,j}\mu(x_i,z_j)H_{m_1}\left(\frac{x-x_i}{h_{n,1}}\right)H_{m_2}\left(\frac{z-z_j}{h_{n,2}}\right)\\
&\,\,-\frac{1}{h_{n,1}h_{n,2}}\iint \mu(u,v)H_{m_1}\left(\frac{x-u}{h_{n,1}}\right)H_{m_2}\left(\frac{z-v}{h_{n,2}}\right)\mathrm{d}u\mathrm{d}v.
\end{split}
\label{mua}
\end{equation}
The first term on the right hand of (\ref{mua}) is the Riemann finite sum of $(h_{n,1}h_{n,2})^{-1}\mu(u,v)$ $H_{m_1}\{h_{n,1}^{-1}(x-u)\}H_{m_2}\{h_{n,2}^{-1}(z-v)\}$ on the grid while the second term is the integral of the same function, and $\mu_0(x,z)$ calculates the difference between the two terms.
$\mu_0(x,z)$ is not random and Lemma~\ref{lem4} shows that $\mu_0(x,z)=O\left\{\max\left (n_1^{-2}h_{n,1}^{-2}, n_2^{-2}h_{n,2}^{-2}\right)\right\}$.  Now (\ref{kernelest2}) becomes
\begin{align}
\textrm{E}\left\{\hat{\mu}(x,z)\right\}& = \frac{1}{h_{n,1}h_{n,2}}\iint \mu(u,v)H_{m_1}\left(\frac{x-u}{h_{n,1}}\right)H_{m_2}\left(\frac{z-v}{h_{n,2}}\right)\mathrm{d}u\mathrm{d}v +\mu_0(x,z) +O(\xi)\nonumber\\
&= \iint\mu(x-h_{n,1}u,z-h_{n,2}v)H_{m_1}(u)H_{m_2}(v)\mathrm{d}u\mathrm{d}v+\mu_0(x,z)+O(\xi).
\label{mean1}
\end{align}
For the double integral in (\ref{mean1}),  we first take the Taylor expansion of $\mu(x-h_{n_1}u, z-h_{n_2}v)$ at $(x,z)$ until the $2m_1$th partial derivative with respect to $x$ and the $2m_2$th  partial derivative with respect to $z$, and then we cancel out those integrals that vanish by Lemma \ref{H_m}.  It follows that explicit expressions for the asymptotic mean  can be attained
\begin{align*}\label{bias}
\textrm{E}\left\{\hat{\mu}(x,z)\right\}-\mu(x,z)-\mu_0(x,z)
& =  (-1)^{m_1+1}h_{n,1}^{2m_1}\frac{\partial^{2m_1}}{\partial x^{2m_1}}\mu(x,z) + (-1)^{m_2+1} h_{n,2}^{2m_2}\frac{\partial^{2m_2}}{\partial z^{2m_2}}\mu(x,z)\\
& +o(h_{n,1}^{2m_1})+o(h_{n,2}^{2m_2})+O(\xi).
\end{align*}
For any two random variables $X$ and $Y$, if $\textrm{var}(Y) = o\{\textrm{var}(X)\}$, then $\textrm{var}(X+Y) = \textrm{var}(X) + o\{\textrm{var}(X)\}$.
Hence, by letting $X=\mu^{\ast}(x,z)$ and $Y = \hat{\mu}(x,z)-\mu^{\ast}(x,z)$, we can obtain by Proposition~\ref{prop1} that
\begin{align*}
\textrm{var}\{\hat{\mu}(x,z)\} = (nh_n)^{-1} \sigma^2(x,z)\int
H_{m_1}^2(u)\mathrm{d}u \int
H_{m_2}^2(v)\mathrm{d}v+o\{(nh_n)\inv\}.
\end{align*}
To get optimal rates of convergence, let $h_{n,1}^{2m_1}/h_{n,2}^{2m_2}$ and $h_{n,1}^{4m_1}/
(nh_n)^{-1}$ converge to some constants, repsectively. Then we have
\begin{align*}
h_{n,1} \sim  h_1n^{-{m_2}/ {m_3} },
h_{n,2} \sim  h_2n^{-{m_1}/{m_3} }
\end{align*}
for some positive constants $h_1$ and $h_2$. (Recall that $m_3 = 4m_1 m_2 + m_1 + m_2$.)   We need to
choose $K_1,K_2$ so that $\max\{(K_1h_{n,1})^{-2},(K_2h_{n,2})^{-2}\}=o(h_{n,1}^{2m_1})$.
Hence, $K_1 \sim C_1 n^{\tau_1}$ for some
positive constant $C_1$ and $\tau_1 > (m_1m_2+m_2)/m_3$. Similarly, $K_2 \sim
C_2n^{\tau_2}$ for some positive constant $C_2$ and
$\tau_2 > (m_1m_2+m_1)/m_3$. It is easy to verify that $\max\left (n_1^{-2}h_{n,1}^{-2}, n_2^{-2}h_{n,2}^{-2}\right) = o(h_{n,1}^{2m_1})$.

\begin{lem}\label{lem3}
Let $G(x)$ be a real function in $[0,1]$ with a continuous second derivative. Let $x_i = (i-1/2)/n$ for $i=1,\dots, n$.  Assume $ h = o(1), (nh^2)^{-1} = o(1)$ as $n$ goes to infinity. Then
$$ \left|\frac{1}{h}\int_0^1 H_m \left(\frac{x-u}{h}\right)G(u) du-\frac{1}{nh} \sum_{i=1}^n H_m \left(\frac{x-x_i}{h}\right)G(x_i)\right| =O(n^{-2}h^{-2}),
$$
where $H_m(x)$ is defined in~(\ref{kernel}).
\end{lem}
{\it Proof of Lemma~\ref{lem3}}: First note that $H_m(x)$ is symmetric and  is bounded by 1. Also $H_m(x)$ is infinitely differentiable over $(-\infty,0]$ and all the derivatives are bounded by $m$ over $(-\infty,0]$.  Let $L_i =[(i-1)/n, i/n]$ for $i=1,\dots, n$. Suppose without loss of generality that $ \max_{u\in[0,1]} |G(u)| \leq m$. We have
\begin{equation}
\label{int1}
\begin{split}
&\left|\frac{1}{h}\int_0^1 H_m \left(\frac{x-u}{h}\right)G(u)du-\frac{1}{nh} \sum_{i=1}^n H_m\left(\frac{x-x_i}{h}\right)G(x_i)\right|\\
 \leq& \sum_{i=1}^n\left|\frac{1}{h}\int_{L_i}\left\{H_m\left(\frac{x-u}{h}\right)G(u)-H_m\left(\frac{x-x_i}{h}\right)G(x_i)\right\}du \right |,
\end{split}
\end{equation}
and
\begin{equation}
\label{int2}
\begin{split}
&\left|\frac{1}{h}\int_{L_i}\left\{H_m \left(\frac{x-u}{h}\right)G(u)-H_m \left(\frac{x-x_i}{h}\right)G(x_i)\right\}du\right| \\
\leq &\left|\frac{G(x_i)}{h}\int_{L_i}\left\{H_m \left(\frac{x-u}{h}\right)-H_m\left(\frac{x-x_i}{h}\right)\right\}du\right|+\left|\frac{1}{h}H_m \left(\frac{x-x_i}{h}\right) \int_{L_i} \left\{G(u)-G(x_i)\right\}du\right|\\
&+\left|\frac{1}{h}\int_{L_i} \left\{H_m \left(\frac{x-u}{h}\right)-H_m\left(\frac{x-x_i}{h}\right)\right\} \left\{G(u)-G(x_i)\right\} du\right|\\
\leq &m\left|\frac{1}{h}\int_{L_i} H_m\left(\frac{x-u}{h}\right)-H_m\left(\frac{x-x_i}{h}\right)du\right|+\frac{1}{h}\left|\int_{L_i} \left\{G(u)-G(x_i)\right\}du\right|+O(n^{-3}h^{-2})\\
\leq &m\left|\frac{1}{h}\int_{L_i}\left\{ H_m\left(\frac{x-u}{h}\right)-H_m\left(\frac{x-x_i}{h}\right)\right\}du\right| +O(n^{-3}h^{-1})+O(n^{-3}h^{-2}).
\end{split}
\end{equation}
In the derivation of~(\ref{int2}), the term $O(n^{-3}h^{-1})$ follows from
$$
\left|G(u)-G(x_i) - (u-x_i) \frac{\partial G}{\partial x}(x_i) \right|\leq \frac{1}{2}(u-x_i)^2\max_{0\leq x\leq 1}\left|\frac{\partial^2 G}{\partial x^2}(x)\right|
$$
and
$$
\left|\int_{L_i} \left\{G(u)-G(x_i)\right\}du\right| =\left|\int_{L_i} \left\{G(u)-G(x_i)- (u-x_i) \frac{\partial G}{\partial x}(x_i) \right\}du\right|;
$$
the term $O(n^{-3}h^{-2})$ follows from
$$
\left|\frac{1}{h} \left\{H_m \left(\frac{x-u}{h}\right)-H_m\left(\frac{x-x_i}{h}\right)\right\} \left\{G(u)-G(x_i)\right\} \right|= O(n^{-2}h^{-2})
$$
since  $|u-x_i| \leq n^{-1}$ when both $u$ and $x_i$ are in $L_i$. Note that we used the equality $\int_{L_i} (u-x_i)du =0$ in the above derivation and we shall use it later as well. Combining~(\ref{int1}) and~(\ref{int2}), we have
\begin{equation}
\label{int3}
\begin{split}
&\left|\frac{1}{h}\int_0^1 H_m \left(\frac{x-u}{h}\right)G(u)du-\frac{1}{nh} \sum_{i=1}^n H_m\left(\frac{x-x_i}{h}\right)G(x_i)\right|\\
 \leq& m\sum_{i=1}^n \left|\frac{1}{h}\int_{L_i}\left\{ H_m\left(\frac{x-u}{h}\right)-H_m\left(\frac{x-x_i}{h}\right)\right\}du\right| +O(n^{-2}h^{-2}).
\end{split}
\end{equation}
For simplicity, denote by $H_m^{(1)}(x)$ and $H_m^{(2)}(x)$ the first and second derivatives of $H_m(x)$, respectively. Similarly, denote  by $H_m^{(1)}(0)$ and $H_m^{(2)}(0)$  the right derivatives of $H_m(x)$ at 0. If $x\in L_i$, then $H_m\left\{h^{-1}(x-u)\right\}- H_m\left\{h^{-1}(x-x_i)\right\} = O(n^{-1}h^{-1})$ and hence
\begin{equation}
\label{int4}
 \left|\frac{1}{h}\int_{L_i}\left\{ H_m\left(\frac{x-u}{h}\right)-H_m\left(\frac{x-x_i}{h}\right)\right\}du\right| = O(n^{-2}h^{-2}),\,\text{if}\,\, x\in L_i.
\end{equation}
If $x<(i-1)/n$, then $x\notin L_i$. Let
\begin{equation*}
\begin{split}
\tilde{H}_m(u,x_i,x,h) = &H_m \left(\frac{x-u}{h}\right)-H_m\left(\frac{x-x_i}{h}\right)-\frac{u-x_i}{h}H_m^{(1)}\left(\frac{x-x_i}{h}\right)\\
&-\frac{(u-x_i)^2}{2h^2}H_m^{(2)}\left(\frac{x-x_i}{h}\right).
\end{split}
\end{equation*}
Then $\tilde{H}_m(u,x_i,x,h) =O(h^{-3}|u-x_i|^3)$. We have
\begin{align}
\label{int5}
&\left|\frac{1}{h}\int_{L_i}\left\{ H_m\left(\frac{x-u}{h}\right)-H_m\left(\frac{x-x_i}{h}\right)\right\}du\right| \nonumber\\
=&\left|\frac{1}{h}\int_{L_i}\left\{ H_m\left(\frac{x-u}{h}\right)-H_m\left(\frac{x-x_i}{h}\right)-\frac{u-x_i}{h}H_m^{(1)}\left(\frac{x-x_i}{h}\right)\right\}du\right| \nonumber\\
\leq&\left|\frac{1}{h}\int_{L_i}\frac{\left(u-x_i\right)^2}{2h^2}H_m^{(2)}\left(\frac{x-x_i}{h}\right)du\right|+ \left|\frac{1}{h}\int_{L_i}\tilde{H}_m(u,x_i,x,h)du\right|\nonumber\\
\leq& \frac{1}{2n^2h^2}\int_{L_i} \frac{1}{h}\left|H_m^{(2)}\left(\frac{x-x_i}{h}\right)\right|du +O(n^{-4}h^{-4}).
\end{align}
We can similarly prove that (\ref{int5}) holds when $x>i/n$. Now with~(\ref{int4}) and~(\ref{int5}),
\begin{equation*}
\begin{split}
&\sum_{i=1}^n \left|\frac{1}{h}\int_{L_i}\left\{ H_m\left(\frac{x-u}{h}\right)-H_m\left(\frac{x-x_i}{h}\right)\right\}du\right|\\
\leq & \frac{1}{2n^2h^2}\int_0^1 \frac{1}{h}\left|H_m^{(2)}\left(\frac{x-x_i}{h}\right)\right|du +O(n^{-3}h^{-4})+O(n^{-2}h^{-2}),
\end{split}
\end{equation*}
which finishes the lemma.

\begin{lem}\label{lem4}
The term $\mu_0(x,z)$ defined in~(\ref{mua}) is $O\left\{\max\left (n_1^{-2}h_{n,1}^{-2}, n_2^{-2}h_{n,2}^{-2}\right)\right\}$.
\end{lem}
{\it Proof of Lemma~\ref{lem4}}:
To simplify notation, let $G_2 (u,z) = h_{n,2}^{-1}\int_0^1H_{m_2}\{h_{n,2}^{-1}(z-v)\} \mu(u, v) dv$ and $G_1(u, z) = (n_2h_{n,2})^{-1} \sum_j H_{m_2}\{h_{n,2}^{-1}(z - z_j)\}\mu(u, z_j) - G_2(u,z)$. Then $G_1$ is $O\{n_2^{-2}h_{n,2}^{-2}\}$ by Lemma~\ref{lem3}. Note that $|\mu_0(x,z)|$ is bounded by the sum of
\begin{equation}
\label{mua_2}
 \left|\frac{1}{n_1h_{n,1}}\sum_i H_{m_1}\left(\frac{x-x_i}{h_{n,1}}\right)G_1(x_i,z)\right|
\end{equation}
and
\begin{equation}
\label{mua_1}
\left|\frac{1}{n_1h_{n,1}}\sum_j H_{m_1}\left(\frac{x-x_i}{h_{n,1}}\right)G_2(x_i,z) - \frac{1}{h_{n,1}} \int H_{m_1}\left(\frac{x-u}{h_{n,1}}\right)G_2(u,z)du \right|.
\end{equation}
Because $G_1$ is $O\left(n_2^{-2}h_{n,2}^{-2}\right)$, (\ref{mua_2}) is also $O\left(n_2^{-2}h_{n,2}^{-2}\right)$. By Theorem 9.1 in the appendix of Durrett (2005), $\partial^2 G_2/\partial u^2$ exists and is equal to $h_{n,2}^{-1}\int_0^1H_{m_2}\{h_{n,2}^{-1}(z-v)\} \partial^2\mu(u,v)/\partial u^2 dv$. Hence $\partial^2 G_2/\partial u^2$ is continuous and bounded. Lemma~\ref{lem3} implies~(\ref{mua_1}) is $O\left(n_1^{-2}h_{n,1}^{-2}\right)$ which finishes our proof.

\textit{Proof of Theorem~\ref{thm2}}: Denote the design points $\{x_i,z_i\}_{i=1}^n$ by $(\underbar{x},\underbar{z})$. Applying  Lemma~\ref{lem1} and the proof of Proposition~\ref{prop1} to the binned data $\tilde{\mathbf{Y}}$  with $n_1,n_2$ replaced by $I_1,I_2$, we obtain
\begin{align}
\textrm{E}\left\{\hat{\mu}(x,z)|(\underbar{x},\underbar{z})\right\}&= (Ih_n)^{-1}\sum_{\kappa,\ell} \textrm{E}\left\{\tilde{y}_{\kappa,\ell}|(\underbar{x},\underbar{z})\right\}G_{\kappa,\ell},\label{mu0}\\
\textrm{var}\left\{\hat{\mu}(x,z)|(\underbar{x},\underbar{z})\right\}&=(Ih_n)^{-2}\sum_{\kappa,\ell} \textrm{var}\left\{\tilde{y}_{\kappa,\ell}|(\underbar{x},\underbar{z})\right\}G_{\kappa,\ell}^2, \label{var0}
\end{align}
where
\[
G_{\kappa,\ell} =H_{m_1}\left(\frac{x-\tilde x_{\kappa}}{h_{n,1}}\right)H_{m_2}\left(\frac{z-\tilde z_{\ell}}{h_{n,2}}\right)+b_{\kappa, \ell}(x,z),
\]
and
$b_{\kappa, \ell}(x,z)$ is defined  similarly to $b_{i,j}(x,z)$ in the proof of Proposition~\ref{prop1}  with also $n_1,n_2$ replaced by $I_1,I_2$.
Let $n_{\kappa,\ell}$ be the number of data points in the $(\kappa,\ell)$th bin. Then
\begin{equation*}
\textrm{var}\left\{\tilde{y}_{\kappa,\ell}|(\underbar{x},\underbar{z})\right\} =n_{\kappa,\ell}^{-2}\sum_{i=1}^{n} \sigma^2(x_i, z_i)\delta_{\{|x_i-\tilde{x}_{\kappa}|\leq (2I_1)^{-1}, |z_i-\tilde{z}_{\ell}|\leq (2I_2)^{-1}\}}.
\end{equation*}
So $\textrm{var}\left\{\sqrt{n_{\kappa,\ell}}\tilde{y}_{\kappa,\ell}|(\underbar{x},\underbar{z})\right\}$ is  a Nadaraya-Watson kernel regression estimator of the conditional variance function $\sigma^2(x,z)$ at $(\tilde{x}_{\kappa}, \tilde{z}_{\ell})$. Similarly, we can show $n_{\kappa,\ell}/(nI^{-1})$ is  a kernel density estimator of $f(x,z)$ at $(\tilde{x}_{\kappa},\tilde{z}_{\ell})$.
By the uniform convergence theory for kernel density estimators and Nadaraya-Watson kernel regression estimators (see, for instance, Hansen (2008)),
\begin{equation}
\label{n_{k,l}}
\sup_{\kappa,\ell}\left|n_{\kappa,\ell}/(nI^{-1}) - f(\tilde{x}_{\kappa},\tilde{z}_{\ell}) \right| = O_p\left\{\sqrt{I\ln n/n} + I^{-2}\right\} = o_p(1),
\end{equation}
and
\begin{equation*}
\sup_{\kappa,\ell}\left| \text{var}\left\{ \sqrt{n_{\kappa,\ell}}\tilde{y}_{\kappa,\ell}|(\underbar{x},\underbar{z})\right\}-\sigma^2(\tilde{x}_{\kappa}, \tilde{z}_{\ell})\right| = O_p\left\{\sqrt{I\ln n/n} + I^{-2}\right\} = o_p(1).
\end{equation*}
It follows by the above two equalities that
\begin{equation}
\label{var1}
\sup_{\kappa,\ell}\left|  \frac{n}{I}\text{var}\left\{\tilde{y}_{\kappa,\ell}|(\underbar{x},\underbar{z})\right\}-\frac{\sigma^2(\tilde{x}_{\kappa}, \tilde{z}_{\ell})}{f(\tilde{x}_{\kappa}, \tilde{z}_{\ell})}\right| = o_p(1).
\end{equation}
By an argument similar to one in the proof of Proposition~\ref{prop1}, for any continuous function $g(x,z)$ over $[0,1]^2$, we can derive that
\begin{equation}
\label{var2}
\frac{1}{Ih_n}\sum_{\kappa,\ell}g(\tilde{x}_{\kappa}, \tilde{z}_{\ell})G^2_{\kappa,\ell} = g(x,z)\int H_{m_1}^2(u)du\int H_{m_2}^2(v)dv + o(1).
\end{equation}
Then by equalities~(\ref{var0}) and~(\ref{var1}),
\begin{equation}
\left|\textrm{var}\left\{\hat{\mu}(x,z)|(\underbar{x},\underbar{z})\right\}-\frac{1}{nh_nIh_n}\sum_{\kappa,\ell}\frac{\sigma^2(\tilde{x}_{\kappa}, \tilde{z}_{\ell})}{f(\tilde{x}_{\kappa}, \tilde{z}_{\ell})}G^2_{\kappa,\ell}\right|= \frac{o_p(1)}{nh_nIh_n}\sum_{\kappa,\ell}G_{\kappa,\ell}^2= o_p\{(nh_n)^{-1}\}. \label{as0}
\end{equation}
By letting $g(x,z) = \sigma^2(x,z)/f(x,z)$ in~(\ref{var2}), we derive from~(\ref{as0}) that
\begin{equation}
\label{as1}
\textrm{var}\left\{\hat{\mu}(x,z)|(\underbar{x},\underbar{z})\right\} = \frac{1}{nh_n}\frac{V(x,z)}{f(x,z)} +o_p\{(nh_n)^{-1}\},
\end{equation}
where $V(x,z)$ is defined in~(\ref{V(x,z)}). We can write $\textrm{E}\left\{\tilde{y}_{\kappa,\ell}|(\underbar{x},\underbar{z})\right\}$ as
\begin{equation*}
\textrm{E}\left\{\tilde{y}_{\kappa,\ell}|(\underbar{x},\underbar{z})\right\} =(n_{\kappa,\ell})^{-1}\sum_{i=1}^{n} \mu(x_i, z_i) \delta_{\{|x_i-\tilde{x}_{\kappa}|\leq (2I_1)^{-1}, |z_i-\tilde{z}_{\ell}|\leq (2I_2)^{-1}\}}.
\end{equation*}
Equality~(\ref{n_{k,l}}) implies each bin is nonempty, so by taking a Taylor expansion of $\mu(x_i, z_j)$ at  $(\tilde{x}_{\kappa},\tilde{z}_{\ell})$ we derive from the above equation that
\begin{equation*}
\sup_{\kappa,\ell} \left|\textrm{E}\left\{\tilde{y}_{\kappa,\ell}|(\underbar{x},\underbar{z})\right\}- \mu(\tilde{x}_{\kappa},\tilde{z}_{\ell}) \right| = O_p(I^{-1/2}).
\end{equation*}
It follows by equality~(\ref{mu0}) that
\begin{align}
\left|\textrm{E}\left\{\hat{\mu}(x,z)|(\underbar{x},\underbar{z})\right\}- \frac{1}{Ih_n}\sum_{\kappa,\ell}\mu(\tilde{x}_{\kappa},\tilde{z}_{\ell})G_{\kappa,\ell}\right| =O_p(I^{-1/2}) \frac{1}{Ih_n}\sum_{\kappa,\ell}|G_{\kappa,\ell}| = O_p(I^{-1/2}).\label{as2}
\end{align}
It is easy to show that
\[
\frac{1}{Ih_n}\sum_{\kappa,\ell}\mu(\tilde{x}_{\kappa},\tilde{z}_{\ell})G_{\kappa,\ell}=\mu(x,z) + n^{-(2m_12m_2)/m_3} \tilde{\mu}(x,z)  + o\left\{n^{-(2m_12m_2)/m_3}\right\},
\]
where $\tilde{\mu}(x,z)$ is defined in~(\ref{tilde_mu}).
In light of equality~(\ref{as2}) and the assumption that $I \sim c_I n^{\tau}$ with $\tau> (4m_1m_2)/m_3$,
\begin{equation}
\label{as3}
\textrm{E}\left\{\hat{\mu}(x,z)|(\underbar{x},\underbar{z})\right\} =\mu(x,z) + n^{-(2m_12m_2)/m_3} \tilde{\mu}(x,z) + o_p\left\{n^{-(2m_12m_2)/m_3}\right\}.
\end{equation}
With~(\ref{as1}) and~(\ref{as3}), we can show that
\begin{equation}
\label{as4}
n^{(2m_12m_2)/m_3}\left[\hat{\mu}(x,z) -\textrm{E}\left\{\hat{\mu}(x,z)|(\underbar{x},\underbar{z})\right\} \right]\Rightarrow N\left\{0, V(x,z)/f(x,z)\right\}
\end{equation}
in distribution and
\begin{equation}
\label{as5}
n^{(2m_12m_2)/m_3}\left[\textrm{E}\left\{\hat{\mu}(x,z)|(\underbar{x},\underbar{z})\right\}-\mu(x, z) \right] = \tilde{\mu}(x,z) + o_p(1).
\end{equation}
Equalities~(\ref{as4}) and (\ref{as5}) together prove the theorem.
\section*{References}
\begin{description}






\item[]
\textsc{Claeskens, G., Krivobokova, T.}, and \textsc{Opsomer, J. D.} (2009),
``Asymptotic properties of penalized spline estimators,"
\textit{Biometrika}, 96, 529-544.

\item[]
\textsc{Currie, I.D., Durban, M.} and \textsc{Eilers, P.H.C.} (2006),
``Generalized linear array models with applications to multidimensional
smoothing,''
\textit{J. R. Statist. Soc. B}, 68, 259-280.

\item[]
\textsc{Di, C., Crainiceanu, C. M., Caffo, B. S.,} and \textsc{Punjabi, N.} (2009),
 ``Multilevel functional principal component analysis,''
 {\it Ann. Appl. Statist.}, 3, 458-488.

\item[]
\textsc{Dierckx, P.} (1982),
``A fast algorithm for smoothing data on a rectangular grid while using spline functions,''
\textit{SIAM J. Numer. Anal.}, 19, 1286-1304.

\item[]
\textsc{Dierckx, P.} (1995),
\textit{Curve and Surface Fitting with Splines},
Clarendon Press, Oxford.

 \item[]
 \textsc{Durrett, R.} (2005),
\textit{Probability: Theory and Examples}, Third Edition,
Thomson.

\item[]
\textsc{Eilers, P.H.C., Currie I.D.} and \textsc{Durban M.} (2006),
``Fast and compact smoothing on large multidimensional
grids,"
\textit{Comput. Statist. Data Anal.},
 50, 61-76.

\item[]
\textsc{Eilers, P.H.C.} and \textsc{Goeman, J.J.} (2004),
``Enhancing scatterplots with smoothed densities,"
\textit{Bioinformatics}, 20, 623-628.

\item[]
\textsc{Eilers, P.H.C.} and \textsc{Marx, B.D.} (1996),
``Flexbile smoothing with B-splines and penalties (with Discussion),"
\textit{Statist. Sci.}, 11, 89-121.

\item[]
\textsc{Eilers, P.H.C.} and \textsc{Marx, B.D.} (2003),
``Mulitvariate calibration with temperature interaction using
 two-dimensional penalized signal regression,"
\textit{Chemometrics and Intelligent Laboratory Systems},
 66, 159-174.

\item[]
\textsc{Gradshteyn, I.S.} and \textsc{Ryzhik, I.M.}(2007),
\textit{Table of Integrals, Series, and Products},
New York: Academic Press.

\item[]
\textsc{Greven, S., Crainiceanu, C., Caffo, B.} and \textsc{Reich, D.} (2010),
``Longitudinal functional principal component,''
\textit{Electronic J. Statist.}, 4, 1022-1054.

\item[]
\textsc{Gu, C.} (2002),
\textit{Smoothing Spline ANOVA Models},
New York: Springer.

\item[]
\textsc{Hall, P., M\" uller, H.G.,} and \textsc{Wang, J.L.} (2006),
``Properties of principal component methods for functional and longitudinal data analysis,"
\textit{Ann. Statist.}, 34, 1493-1517.

\item[]
\textsc{Hansen, B.E.} (2008),
``Uniform convergence rates for kernel estimation with dependent data,"
\textit{Econometric Theory}, 24, 726-748.

\item[]
\textsc{Hastie, T.} and \textsc{Tibshirani, R.}(1993),
``Varying-coefficients models,"
 \textit{J. R. Statist. Soc. B}, 55, 757-796.


 \item[]
 \textsc{Laub, A.J.}(2005),
\textit{Matrix Analysis for Scientists and Engineers},
SIAM.

\item[]
\textsc{Kauermann, G., Krivobokova, T.} and \textsc{Fahrmeir, L.} (2009),
``Some asymptotic results on generalized penalized spline smoothing,''
\textit{J. R. Statist. Soc. B}, 71, 487-503.

\item[]
\textsc{Li,Y.} and \textsc{Ruppert D.} (2008),
``On the asymptotics of penalized splines,"
\textit{Biometrika}, 95, 415-436.


\item[]
\textsc{Marx, B.D.} and \textsc{Eilers, P.H.C.} (2005),
``Multdimensional Penalized Signal Regression,"
\textit{Technometrics}, 47, 13-22.

\item[]
\textsc{Opsomer, J.D.} and \textsc{Hall, P.} (2005),
``Theory for penalised spline regression,"
\textit{Biometrika}, 95, 417-436.

\item[]
\textsc{Ruppert, D.} (2002),
``Selecting the number of knots for penalized splines,"
\textit{J. Comput. Graph. Statist.}, 1, 735-757.

\item[]
\textsc{Ruppert, D., Wand, M.P.} and \textsc{Carroll, R.J.} (2003),
\textit{Semiparametric Regression},
 Cambridge: Cambridge University Press.


\item[]
\textsc{Seber, G.A.F.} (2007),
\textit{ A Matrix Handbook for Statisticians},
New Jersey: Wiley-Inter\-sci\-ence.

\item[]
\textsc{Staniswalis, J.G.} and \textsc{Lee, J.J.} (1998),
``Nonparametric regression analysis of longitudinal data,"
\textit{J. Amer. Statist. Assoc.}, 93, 1403-1418.

\item[]
\textsc{Stone, C.J.} (1980),
``Optimial rates of convergence for nonparametric estimators,''
\textit{Ann. Statist.}, 8, 1348-1360.


\item[]
\textsc{Wand, M.P.} and \textsc{Jones, M.C.} (1995),
\textit{Kernel Smoothing},
London: Chapman \&Hall.

\item[]
\textsc{Wang, X., Shen, J.} and \textsc{Ruppert, D.} (2011),
``Local Asymptotics of P-spline Smoothing,''
\textit{Electronic J. Statist.}, 4, 1-17.


\item[]
\textsc{Wood, S.N.} (2003),
``Thin plate regression splines,"
\textit{J. R. Statist. Soc. B}, 65, 95-114.

\item[]
\textsc{Wood, S.N.} (2006),
\textit{Generalized additive models: an introduction with R},
London: Chapman \&Hall.

\item[]
\textsc{Xiao, L., Li, Y., Apanasovich, T.V.} and \textsc{Ruppert, D.} (2011),
``Local asymptotics of P-splines,"
available at \url{http://arxiv.org/abs/1201.0708v3}.

\item[]
\textsc{Yao, F.}  and \textsc{Lee C.M.} (2006),
``Penalized spline models for functional principal component analysis,"
\textit{J. R. Statist. Soc. B}, 68, 3-25.

\item[]
\textsc{Yao, F., M\" uller, H.G.,} and \textsc{Wang, J.L.} (2005),
``Functional data analysis for sparse longitudinal data,"
\textit{J. Amer. Statist. Assoc.}, 100, 577-590.

\item[]
\textsc{Zipunnikov, V., Caffo, B. S., Crainiceanu, C. M., Yousem D.M., Davatzikos, C.,} and \textsc{Schwartz, B.S.} (2011),
``Multilevel functional principal component analysis for high-dimensional data,''
\textit{J. Comput. Graph. Statist.}, 20(4), 852-873.


\end{description}

\end{document}